\documentclass[
aps,
pra,
reprint,
superscriptaddress,
nofootinbib
]{revtex4-2}

\usepackage{amsmath}
\usepackage{amssymb}
\usepackage{amsfonts}
\usepackage{bm}
\usepackage{mathtools}
\usepackage{graphicx}
\usepackage{booktabs}
\usepackage{multirow}
\usepackage{array}
\usepackage{xcolor}
\usepackage{url}
\usepackage{tikz}
\usetikzlibrary{
    arrows.meta,
    calc,
    decorations.pathmorphing,
    positioning,
    shadings
}

\allowdisplaybreaks[4]

\newcommand{\Tr}{\operatorname{Tr}}

\newcommand{\ket}[1]{\left|#1\right\rangle}
\newcommand{\bra}[1]{\left\langle#1\right|}

\begin{document}

\title{Quantumness of Entanglement in Schwarzschild de Sitter Spacetime}

\author{Sumeng Wang}
\affiliation{
School of Mathematics and Statistics,
Hainan Normal University,
Haikou 571158, China
}

\author{Xiaofen Huang}
\email{huangxf1206@163.com}
\affiliation{
School of Mathematics and Statistics,
Hainan Normal University,
Haikou 571158, China
}

\begin{abstract}

We investigate genuine multipartite entanglement and genuine
tripartite nonlocality of a noisy generalized GHZ state in
Schwarzschild--de Sitter spacetime. The genuine multipartite
concurrence and optimized Svetlichny parameter are employed to
characterize these two quantum correlations. It is shown that the
black hole and cosmological event horizons redistribute genuine
multipartite entanglement differently among the physically accessible
and inaccessible reduced states, while Svetlichny violation occurs
only in the physically accessible state. We find that genuine
multipartite entanglement can survive in the physically inaccessible
states even when genuine tripartite nonlocality is absent. In addition,
initial white noise suppresses both correlations, with Svetlichny
nonlocality surviving over a narrower parameter region. Furthermore,
in the Nariai limit, under the normalization adopted here, the
correlations become concentrated in the physically accessible state.
These results demonstrate the different robustness of genuine
multipartite entanglement and genuine tripartite nonlocality under
horizon effects and initial noise.

\end{abstract}

\maketitle

\section{Introduction}
\label{sec:introduction}

Quantum information theory provides a fundamental framework for
understanding how information is encoded, transmitted, and processed
in quantum systems. Its extension to relativistic and gravitational
settings is of particular interest because the notions of particles,
vacuum states, and quantum correlations depend on the observer and
the spacetime geometry. The study of quantum information in curved
spacetime is therefore relevant not only to fundamental questions at
the interface of quantum theory and gravitation, but also to the
behavior of quantum resources when relativistic effects cannot be
neglected. Hawking's discovery of particle creation by black holes
revealed the thermal character of quantum fields near a black-hole
event horizon \cite{Hawking1975}, while Gibbons and Hawking showed
that cosmological event horizons possess an analogous thermodynamic
structure \cite{Gibbons1977}. These developments established a deep
connection among quantum field theory, gravitation, and
thermodynamics and laid the foundation for quantum field theory in
curved spacetime \cite{Birrell1982,Wald1994}.

The observer dependence of particle content has important
consequences for quantum information. In noninertial and curved
spacetimes, correlations initially encoded in global field modes can
be redistributed when part of the field becomes inaccessible to a
given observer. This phenomenon was first explored extensively in
uniformly accelerated frames and black hole backgrounds
\cite{Unruh1976,FuentesSchuller2005,Alsing2006,Bruschi2010,
MartinMartinez2010}. Subsequent studies showed that the behavior of
entanglement depends strongly on the field statistics, the initial
state, and the observer's motion
\cite{Adesso2007,PanJing2008,MartinMartinezLeon2010,
MartinMartinezFuentes2011,Montero2011,XiaoFang2011,Khan2014}.
Relativistic effects on multipartite fermionic systems have also been
investigated, revealing that multipartite correlations provide
features that cannot be inferred directly from their bipartite
counterparts \cite{WangJing2011,Qiang2018}. More recent work in
Schwarzschild spacetime has further demonstrated that Hawking
radiation can modify physically accessible and inaccessible
correlations and affect quantum-information protocols in nontrivial
ways
\cite{Elghaayda2024Correlations,Wu2023Teleportation,
WuTeng2024Accessible}.

Most of these investigations have focused on geometries characterized
by a single event horizon. Schwarzschild--de Sitter (SdS) spacetime
provides a qualitatively different setting because it contains both a
black hole event horizon (BEH) and a cosmological event horizon (CEH)
\cite{Kottler1918,Nariai1951}. The thermodynamic properties of de
Sitter black holes and the roles of the two horizons have been
examined from several perspectives
\cite{Gomberoff2003,Sekiwa2006,Bhattacharya2018}. In general, the BEH
and CEH possess different surface gravities and therefore correspond
to distinct thermal sectors. Moreover, their coordinate
singularities cannot be removed simultaneously by a single Kruskal
extension. A thermally opaque membrane introduced in the static
region provides a useful way to separate the two sectors and to treat
the corresponding fermionic transformations independently
\cite{Liu2023MultiHorizon}. This two horizon structure makes SdS
spacetime a natural arena for investigating how quantum correlations
are redistributed when different parts of a multipartite system are
affected by different horizon transformations.

Quantum correlations in SdS and related multiple-horizon spacetimes
have attracted increasing attention. Entanglement degradation in
multi event horizon geometries has been analyzed for quantum fields
\cite{BhattacharyaJoshi2022}, while fermionic entanglement, mutual
information, and other quantum properties have been investigated in
SdS spacetime \cite{Liu2023MultiHorizon}. Quantum steering in such
backgrounds has also been studied
\cite{Wu2024Steering}. More recently, genuine multipartite
entanglement of Dirac fields in SdS spacetime has been considered for
general \(N\) partite systems
\cite{Wu2024GME,HuangLiu2025GME}. These developments indicate that
the presence of two event horizons can lead to a richer distribution
of quantum correlations than in single-horizon spacetimes.

Multipartite entanglement is particularly important because it
represents correlations that cannot, in general, be reduced to a
collection of bipartite ones. The
Greenberger Horne Zeilinger state provides a paradigmatic example
of genuine tripartite entanglement
\cite{GHZ1989}, while the classification of multipartite quantum
states demonstrates the existence of inequivalent entanglement
structures with no direct bipartite analogue
\cite{Dur2000,Acin2000}. Multipartite entanglement in relativistic
settings has consequently been studied in a variety of scenarios
\cite{WangJing2011,Qiang2018,WuTeng2024Accessible}.
Recent investigations based on Unruh--DeWitt detectors have further
shown that relativistic motion can give rise to nontrivial changes in
multipartite entanglement
\cite{LiShangWu2025,LiYangXu2026}. For multiqubit X states, genuine
multipartite entanglement can be characterized analytically by the
genuine multipartite concurrence
\cite{Ma2011,Hashemi2012}, making this measure particularly suitable
for the reduced states generated by fermionic horizon
transformations.

Entanglement, however, does not exhaust the nonclassical correlation
structure of a multipartite quantum system. The question of whether
quantum correlations can be reproduced by a local hidden-variable
description lies at the heart of the foundations of quantum theory
\cite{EPR1935,Brunner2014}. Bell nonlocality also provides an
important operational resource for device independent quantum
cryptography, certified randomness generation, and randomness
amplification
\cite{Acin2007,Pironio2010,ColbeckRenner2012}. For tripartite
systems, genuine multipartite nonlocality can be investigated through
the Svetlichny inequality, which distinguishes genuinely
three-party nonlocal correlations from those compatible with hybrid
local and nonlocal hidden-variable models
\cite{Svetlichny1987}. The relation between entanglement and
Svetlichny nonlocality has been examined for three qubit systems and
in relativistic settings
\cite{AjoyRungta2010,Tian2013,Wang2020GenuineNonlocality}.
Studies in curved spacetime have further addressed genuine
tripartite nonlocality and its relation to physically accessible and
inaccessible correlations
\cite{WuZeng2022Genuine,ZhangWangFei2023}. Genuine four partite Bell
nonlocality has also been considered in curved spacetime
\cite{Mi2025BellNonlocality}, while recent work has extended the
analysis of relativistic nonlocal correlations to quantum networks
\cite{LiWangTongWu2026Network}.

Despite these developments, several questions remain open.
Most existing studies of SdS spacetime have focused on individual
quantum correlation measures, whereas a direct comparison between
genuine multipartite entanglement and genuine tripartite nonlocality
for a noisy tripartite system subjected simultaneously to the BEH and
CEH transformations remains limited. In particular, it is not yet
clear how these two different forms of multipartite quantum
correlation are distributed among the physically accessible and
inaccessible sectors, or whether they possess comparable robustness
against initial noise. In addition, the combined influence of the
spacetime geometry, the initial state preparation, and the field
frequency has not been systematically examined for both quantities
within the same two-horizon framework. Clarifying these issues is
important for understanding which aspects of multipartite quantum
correlations remain operationally relevant in curved spacetimes and
how entanglement and nonlocality respond differently to gravitational
effects.

Motivated by these questions, the present work investigates a noisy
generalized GHZ state in Schwarzschild--de Sitter spacetime.
Alice's mode is associated with the BEH, Charlie's mode with the CEH,
while Bob retains a reference mode in the static region. The
fermionic transformations associated with the two horizons map the
initial three mode state onto a five mode system. By tracing over the
appropriate modes, four three qubit reduced states are obtained: the
physically accessible state \(\rho_{ABC}\) and the physically
inaccessible reduced states
\(\rho_{ABR}\), \(\rho_{LBC}\), and \(\rho_{LBR}\).
The genuine multipartite concurrence is employed to characterize
genuine multipartite entanglement, while the optimized Svetlichny
parameter is used to examine genuine tripartite nonlocality.
Analytical expressions are derived for the four reduced states, and
the effects of the spacetime parameter, initial white noise,
generalized GHZ parameter, and field frequency are investigated.
This setting provides a unified framework for comparing genuine
multipartite entanglement and genuine tripartite nonlocality in a
two horizon spacetime.

The remainder of this paper is organized as follows.
Sec.~\ref{sec:SdS} introduces the Schwarzschild--de Sitter geometry
and the fermionic mode transformations associated with the
black hole and cosmological event horizons.
Sec.~\ref{sec:resources} constructs the noisy generalized GHZ state,
derives the four reduced states, and introduces the genuine
multipartite concurrence and optimized Svetlichny parameter.
Sec.~\ref{sec:results} presents the numerical analysis of the
multipartite correlations under variations of the spacetime and
initial-state parameters.
Finally, Sec.~\ref{sec:conclusion} summarizes the main conclusions.


\section{Dirac fields in Schwarzschild de Sitter spacetime}
\label{sec:SdS}

We briefly introduce the geometric and field theoretic ingredients
required for the subsequent analysis of tripartite quantum resources
in Schwarzschild de Sitter spacetime. Throughout this work, natural
units
\(
G=c=\hbar=k_{\mathrm{B}}=1
\)
are adopted.


\subsection{Schwarzschild de Sitter spacetime}

The Schwarzschild de Sitter spacetime describes a static,
spherically symmetric black hole in the presence of a positive
cosmological constant
\cite{Liu2023MultiHorizon,BhattacharyaJoshi2022,
Bhattacharya2018}.
Its line element is
\begin{equation}
\begin{aligned}
ds^{2}
={}&
-f(r)\,dt^{2}
+\frac{dr^{2}}{f(r)}
+r^{2}
\left(
d\theta^{2}
+\sin^{2}\theta\,d\phi^{2}
\right),
\end{aligned}
\label{eq:SdS_metric}
\end{equation}
where
\begin{equation}
f(r)
=
1-\frac{2M}{r}
-\frac{\Lambda r^{2}}{3}.
\label{eq:SdS_function}
\end{equation}
Here, \(M\) denotes the black hole mass and
\(\Lambda>0\) is the cosmological constant.

For \(0<3M\sqrt{\Lambda}<1\), 
the SdS spacetime possesses two positive roots of \(f(r)=0\),
corresponding to the black hole event horizon (BEH) and the
cosmological event horizon (CEH)
\cite{BhattacharyaJoshi2022,Liu2023MultiHorizon}.
Their radii are
\begin{equation}
\begin{aligned}
r_{\mathcal{H}}
&=
\frac{2}{\sqrt{\Lambda}}
\cos
\left[
\frac{
\arccos(-3M\sqrt{\Lambda})+4\pi
}{3}
\right],
\\[4pt]
r_{\mathcal{C}}
&=
\frac{2}{\sqrt{\Lambda}}
\cos
\left[
\frac{
\arccos(-3M\sqrt{\Lambda})
}{3}
\right].
\end{aligned}
\label{eq:horizon_radii}
\end{equation}

These two positive roots satisfy
\(r_{\mathcal{H}}<r_{\mathcal{C}}\). The static region relevant to
the present analysis is therefore
\(r_{\mathcal{H}}<r<r_{\mathcal{C}}\).

For later convenience, the positive magnitudes of the surface
gravities at the two event horizons are used
\cite{BhattacharyaJoshi2022,Liu2023MultiHorizon}.
They are given by
\begin{equation}
\begin{aligned}
\kappa_{\mathcal{H}}
&=
\frac{
\Lambda
(2r_{\mathcal{H}}+r_{\mathcal{C}})
(r_{\mathcal{C}}-r_{\mathcal{H}})
}{
6r_{\mathcal{H}}
},
\\[4pt]
\kappa_{\mathcal{C}}
&=
\frac{
\Lambda
(2r_{\mathcal{C}}+r_{\mathcal{H}})
(r_{\mathcal{C}}-r_{\mathcal{H}})
}{
6r_{\mathcal{C}}
}.
\end{aligned}
\label{eq:kappa_HC}
\end{equation}
The corresponding horizon temperatures are
\(T_{\mathcal{H}}=\kappa_{\mathcal{H}}/(2\pi)\) and
\(T_{\mathcal{C}}=\kappa_{\mathcal{C}}/(2\pi)\).
Since the two surface gravities are generally unequal,
\(\kappa_{\mathcal{H}}\neq\kappa_{\mathcal{C}}\), the BEH and CEH
generate distinct thermal effects. This difference will subsequently
lead to different fermionic mode transformations at the two event
horizons.

An important limiting configuration is the Nariai limit
\cite{Nariai1951,Liu2023MultiHorizon}, for which
\(3M\sqrt{\Lambda}\to1\) and the two event horizons coincide,
\(r_{\mathcal{H}}=r_{\mathcal{C}}=1/\sqrt{\Lambda}\).
Within the normalization adopted here,
\(\kappa_{\mathcal{H}}\) and \(\kappa_{\mathcal{C}}\) both approach
zero in this limit. Consequently, the horizon induced thermal
effects are suppressed for fermionic modes of finite frequency.

Unlike a spacetime containing only one event horizon, the coordinate
singularities associated with the BEH and CEH cannot be removed
simultaneously by a single Kruskal extension
\cite{Bhattacharya2018,BhattacharyaJoshi2022,
Liu2023MultiHorizon}.
Following the multi event horizon construction, the two thermal
sectors can be treated independently by introducing a thermally
opaque membrane in the static region between the horizons
\cite{Gomberoff2003,Sekiwa2006,
Bhattacharya2018,BhattacharyaJoshi2022,
Liu2023MultiHorizon}.
This separation prevents modes localized near one event horizon
from entering the thermal sector associated with the other horizon
and allows the two fermionic transformations to be considered
independently.


\subsection{Fermionic vacuum structure at the two event horizons}

We now introduce the fermionic mode transformations required to
construct the tripartite state in the following section. The
quantization of Dirac fields in noninertial and black hole
backgrounds shows that local and global fermionic modes are related
through Bogoliubov transformations
\cite{FuentesSchuller2005,Alsing2006,
MartinMartinez2010,Liu2023MultiHorizon}.
Because the BEH and CEH possess different surface gravities, their
associated transformations must be treated independently.

For a fermionic mode associated with the black hole event horizon,
the Kruskal vacuum can be written as
\cite{Alsing2006,MartinMartinez2010,
Liu2023MultiHorizon}
\begin{equation}
\begin{aligned}
\ket{0}_{\kappa_{\mathcal{H}}}
={}&
\mu_{\mathcal{H}}\ket{0_A0_L}
+
\nu_{\mathcal{H}}\ket{1_A1_L},
\end{aligned}
\label{eq:vacuum_H}
\end{equation}
where \(A\) denotes the physically accessible mode on the black hole
side of the membrane and \(L\) denotes its causally inaccessible
partner mode. The corresponding excited state is
\begin{equation}
\ket{1}_{\kappa_{\mathcal{H}}}
=
\ket{1_A0_L}.
\label{eq:excited_H}
\end{equation}
The corresponding Bogoliubov coefficients are given by
\begin{equation}
\begin{aligned}
\mu_{\mathcal{H}}
&=
\left[
1+
\exp
\left(
-\frac{2\pi\omega_A}{\kappa_{\mathcal{H}}}
\right)
\right]^{-1/2},
\\[4pt]
\nu_{\mathcal{H}}
&=
\left[
1+
\exp
\left(
\frac{2\pi\omega_A}{\kappa_{\mathcal{H}}}
\right)
\right]^{-1/2}.
\end{aligned}
\label{eq:mu_nu_H}
\end{equation}
which satisfy
\(\mu_{\mathcal{H}}^{2}+\nu_{\mathcal{H}}^{2}=1\).

Similarly, for a fermionic mode associated with the cosmological
event horizon, the Kruskal vacuum is
\cite{Liu2023MultiHorizon}
\begin{equation}
\begin{aligned}
\ket{0}_{\kappa_{\mathcal{C}}}
={}&
\mu_{\mathcal{C}}\ket{0_C0_R}
+
\nu_{\mathcal{C}}\ket{1_C1_R},
\end{aligned}
\label{eq:vacuum_C}
\end{equation}
where \(C\) denotes the physically accessible mode on the
cosmological horizon side of the membrane and \(R\) is its causally
inaccessible partner mode. The corresponding excited state is
\begin{equation}
\ket{1}_{\kappa_{\mathcal{C}}}
=
\ket{1_C0_R}.
\label{eq:excited_C}
\end{equation}
The associated Bogoliubov coefficients are
\begin{equation}
\begin{aligned}
\mu_{\mathcal{C}}
&=
\left[
1+
\exp
\left(
-\frac{2\pi\omega_C}{\kappa_{\mathcal{C}}}
\right)
\right]^{-1/2},
\\[4pt]
\nu_{\mathcal{C}}
&=
\left[
1+
\exp
\left(
\frac{2\pi\omega_C}{\kappa_{\mathcal{C}}}
\right)
\right]^{-1/2}.
\end{aligned}
\label{eq:mu_nu_C}
\end{equation}
which satisfy
\(\mu_{\mathcal{C}}^{2}+\nu_{\mathcal{C}}^{2}=1\).

The two sets of Bogoliubov coefficients,
\((\mu_{\mathcal{H}},\nu_{\mathcal{H}})\) and
\((\mu_{\mathcal{C}},\nu_{\mathcal{C}})\), are generally different
because they are governed by the independent surface gravities
\(\kappa_{\mathcal{H}}\) and \(\kappa_{\mathcal{C}}\). This distinction
is the essential feature required in the following analysis, since
the two event horizons act independently on different components of
the tripartite state.

For finite mode frequencies, the Nariai limit gives
\(\mu_{\mathcal{H}},\mu_{\mathcal{C}}\to1\) and
\(\nu_{\mathcal{H}},\nu_{\mathcal{C}}\to0\) within the normalization
adopted in this work. Thus, the fermionic mixing induced by the two
horizons is suppressed in this limit.

\begin{figure}[!t]
\centering
\resizebox{0.98\columnwidth}{!}{%
\begin{tikzpicture}[
    x=1cm,
    y=1cm,
    >=Latex,
    line cap=round,
    line join=round,
    font=\small,
    decorate
]


\definecolor{bgblue}{RGB}{225,238,248}
\definecolor{behcol}{RGB}{245,185,30}
\definecolor{cehcol}{RGB}{70,200,255}
\definecolor{Acol}{RGB}{233,148,55}
\definecolor{Bcol}{RGB}{112,170,110}
\definecolor{Ccol}{RGB}{90,190,240}
\definecolor{Lcol}{RGB}{252,241,170}
\definecolor{Rcol}{RGB}{225,242,252}
\definecolor{arrowpurple}{RGB}{198,145,222}
\definecolor{textdark}{RGB}{35,45,65}
\definecolor{wavecol}{RGB}{150,165,180}

\fill[bgblue] (0,0) rectangle (11.8,7.1);

\coordinate (BH) at (1.45,3.05);

\shade[ball color=black!88] (BH) circle (1.05);
\draw[orange!85!black, line width=0.7pt] (BH) circle (1.05);

\fill[white,opacity=0.22]
($(BH)+(-0.34,0.33)$) ellipse (0.28 and 0.11);
\fill[white,opacity=0.08]
($(BH)+(-0.08,0.08)$) ellipse (0.17 and 0.06);

\node[
    text=textdark,
    font=\bfseries\large
] at ($(BH)+(0,-1.42)$) {Black hole};


\coordinate (L)  at (3.10,2.15);
\coordinate (A1) at (5.05,2.15);
\coordinate (B1) at (6.45,2.15);
\coordinate (C1) at (7.85,2.15);
\coordinate (R)  at (10.10,2.15);

\coordinate (A0) at (5.05,5.25);
\coordinate (B0) at (6.45,5.25);
\coordinate (C0) at (7.85,5.25);

\coordinate (BEHc) at (4.10,3.18);
\coordinate (CEHc) at (8.95,3.18);


\draw[
    behcol,
    line width=1.2pt
]
(BEHc) ellipse (0.34 and 2.42);

\draw[
    behcol,
    line width=1.2pt
]
($(BEHc)+(0,-2.15)$) -- ($(BEHc)+(0,2.15)$);

\draw[
    cehcol,
    line width=1.2pt
]
(CEHc) ellipse (0.36 and 2.42);

\draw[
    cehcol,
    line width=1.2pt
]
($(CEHc)+(0,-2.15)$) -- ($(CEHc)+(0,2.15)$);

\node[
    text=behcol!85!black,
    font=\Large
] at ($(BEHc)+(-0.08,2.72)$) {$r_{\mathcal{H}}$};

\node[
    text=cehcol!75!black,
    font=\Large
] at ($(CEHc)+(0.08,2.72)$) {$r_{\mathcal{C}}$};

\node[
    text=behcol!85!black,
    font=\bfseries\small,
    align=center
] at ($(BEHc)+(0,-2.88)$) {black hole event horizon};

\node[
    text=cehcol!75!black,
    font=\bfseries\small,
    align=center
] at ($(CEHc)+(0,-2.88)$) {cosmological event horizon};


\shade[ball color=Acol] (A0) circle (0.42);
\shade[ball color=Bcol] (B0) circle (0.42);
\shade[ball color=Ccol] (C0) circle (0.42);

\node[font=\bfseries\Large] at (A0) {$A$};
\node[font=\bfseries\Large] at (B0) {$B$};
\node[font=\bfseries\Large] at (C0) {$C$};

\foreach \yy in {-0.10,0,0.10}{
    \draw[
        wavecol!85,
        line width=0.7pt,
        decoration={
            snake,
            amplitude=1.0pt,
            segment length=5.3pt
        }
    ]
    ($(A0)+(0.43,\yy)$)
    --
    ($(B0)+(-0.43,\yy)$);

    \draw[
        wavecol!85,
        line width=0.7pt,
        decoration={
            snake,
            amplitude=1.0pt,
            segment length=5.3pt
        }
    ]
    ($(B0)+(0.43,\yy)$)
    --
    ($(C0)+(-0.43,\yy)$);
}


\draw[
    arrowpurple,
    fill=arrowpurple!80,
    line width=1.0pt,
    -{Latex[length=8mm,width=6mm]}
]
(6.45,4.15)
--
(6.45,3.35);


\shade[ball color=Lcol] (L) circle (0.40);
\shade[ball color=Acol] (A1) circle (0.42);
\shade[ball color=Bcol] (B1) circle (0.42);
\shade[ball color=Ccol] (C1) circle (0.42);
\shade[ball color=Rcol] (R) circle (0.40);

\node[font=\bfseries\Large] at (L) {$L$};
\node[font=\bfseries\Large] at (A1) {$A$};
\node[font=\bfseries\Large] at (B1) {$B$};
\node[font=\bfseries\Large] at (C1) {$C$};
\node[font=\bfseries\Large] at (R) {$R$};


\foreach \dy in {-0.12,0,0.12}{

    \draw[
        wavecol!90,
        line width=0.75pt,
        decoration={
            snake,
            amplitude=1.0pt,
            segment length=5.0pt
        }
    ]
    ($(L)+(0.40,\dy)$)
    --
    ($(A1)+(-0.42,\dy)$);

    \draw[
        wavecol!90,
        line width=0.75pt,
        decoration={
            snake,
            amplitude=1.0pt,
            segment length=5.0pt
        }
    ]
    ($(A1)+(0.42,\dy)$)
    --
    ($(B1)+(-0.42,\dy)$);

    \draw[
        wavecol!90,
        line width=0.75pt,
        decoration={
            snake,
            amplitude=1.0pt,
            segment length=5.0pt
        }
    ]
    ($(B1)+(0.42,\dy)$)
    --
    ($(C1)+(-0.42,\dy)$);

    \draw[
        wavecol!90,
        line width=0.75pt,
        decoration={
            snake,
            amplitude=1.0pt,
            segment length=5.0pt
        }
    ]
    ($(C1)+(0.42,\dy)$)
    --
    ($(R)+(-0.40,\dy)$);
}

\end{tikzpicture}%
}
\caption{
Schematic illustration of the tripartite configuration in
Schwarzschild de Sitter spacetime. The upper row represents the
initial modes \(A\), \(B\), and \(C\), while the lower row shows the
redistributed modes \(L\), \(A\), \(B\), \(C\), and \(R\) after the
two horizon induced transformations.
}
\label{fig:observer_configuration}
\end{figure}
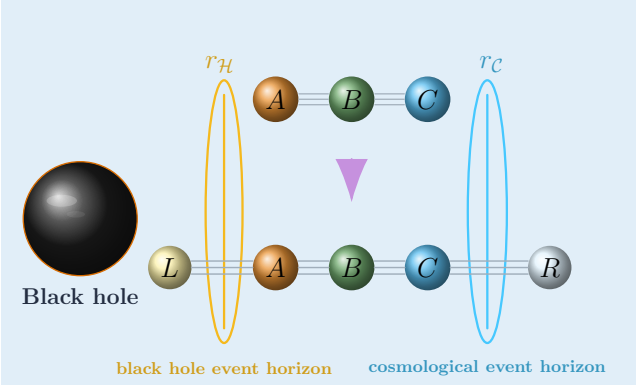

In the tripartite configuration considered below, Alice's mode is
associated with the black hole event horizon, Charlie's mode is
associated with the cosmological event horizon, and Bob retains a
reference mode in the static region. The corresponding mode assignment is illustrated schematically in
Fig.~\ref{fig:observer_configuration}. Here,
\(A\) and \(C\) denote the physically accessible modes associated
with the black hole and cosmological event horizons, respectively,
while \(L\) and \(R\) denote their causally inaccessible partner
modes. Bob's mode \(B\) remains unchanged.

\section{Tripartite quantum resources in Schwarzschild de Sitter spacetime}
\label{sec:resources}

The fermionic transformations associated with the black hole event
horizon and cosmological event horizon, introduced in
Sec.~\ref{sec:SdS}, are now applied to a tripartite mixed state.
Alice's mode is associated with the BEH, Charlie's mode with the CEH,
while Bob retains a reference mode in the static region. The
resulting reduced states are used to characterize genuine
multipartite entanglement and genuine tripartite nonlocality
\cite{WangJing2011,Qiang2018,Tian2013,
Wang2020GenuineNonlocality}.


\subsection{Generalized GHZ state and horizon transformation}
\label{sec:state}

The initial state shared by Alice, Bob, and Charlie is taken to be a
generalized GHZ state subject to white noise,
\begin{equation}
\rho_{ABC}^{(0)}
=
p\ket{\Psi}\bra{\Psi}
+
\frac{1-p}{8}\mathbb{I}_{8},
\label{eq:initial_mixed_GHZ}
\end{equation}
where
\begin{equation}
\ket{\Psi}_{ABC}
=
\alpha\ket{000}_{ABC}
+
\sqrt{1-\alpha^{2}}\ket{111}_{ABC}.
\label{eq:generalized_GHZ}
\end{equation}
Here,
\(0\leq p\leq1\) and \(0\leq\alpha\leq1\).
The parameter \(p\) specifies the weight of the generalized GHZ
component in the initial mixed state, whereas \(\alpha\) determines
the relative amplitudes of the \(\ket{000}\) and \(\ket{111}\)
components. Thus, \(p=1\) corresponds to a pure generalized GHZ
state.

For later convenience, we define
\(\beta=\sqrt{1-\alpha^{2}}\) and
\(q=(1-p)/8\).
The quantity \(q\) denotes the contribution of the white-noise term
to each computational-basis component and is not an independent
state parameter.

To describe the action of the black hole event horizon (BEH) and
cosmological event horizon (CEH) on the complete mixed state, the
fermionic transformations associated with the two horizons are
represented by the local isometries
\(V_{\mathcal H}\) and \(V_{\mathcal C}\), respectively.
Using the input bases
\(\{\ket{0_A},\ket{1_A}\}\) and
\(\{\ket{0_C},\ket{1_C}\}\), together with the ordered output bases
\[
\{
\ket{0_A0_L},
\ket{0_A1_L},
\ket{1_A0_L},
\ket{1_A1_L}
\}
\]
and
\[
\{
\ket{0_C0_R},
\ket{0_C1_R},
\ket{1_C0_R},
\ket{1_C1_R}
\},
\]
the two isometries can be written as
\begin{equation}
V_{\mathcal H}
=
\begin{pmatrix}
\mu_{\mathcal H} & 0 \\
0 & 0 \\
0 & 1 \\
\nu_{\mathcal H} & 0
\end{pmatrix},
\qquad
V_{\mathcal C}
=
\begin{pmatrix}
\mu_{\mathcal C} & 0 \\
0 & 0 \\
0 & 1 \\
\nu_{\mathcal C} & 0
\end{pmatrix}.
\label{eq:local_isometries}
\end{equation}
These matrices reproduce the vacuum and excited state transformations
introduced in Sec.~\ref{sec:SdS}.
Since
\(\mu_{\mathcal H}^{2}+\nu_{\mathcal H}^{2}=1\) and
\(\mu_{\mathcal C}^{2}+\nu_{\mathcal C}^{2}=1\),
they satisfy
\begin{equation}
V_{\mathcal H}^{\dagger}V_{\mathcal H}
=
\mathbb{I}_{A},
\qquad
V_{\mathcal C}^{\dagger}V_{\mathcal C}
=
\mathbb{I}_{C}.
\label{eq:isometry_conditions}
\end{equation}

Bob's mode is unaffected by either horizon transformation.
The total isometry acting on the initial tripartite state is therefore
\begin{equation}
V
=
V_{\mathcal H}
\otimes
\mathbb{I}_{B}
\otimes
V_{\mathcal C}.
\label{eq:total_isometry}
\end{equation}
The transformed density operator can be represented as
\begin{equation}
\rho_{ALBCR}
=
V
\rho_{ABC}^{(0)}
V^{\dagger},
\label{eq:five_mode_density}
\end{equation}
where the five modes are ordered as
\(A\,L\,B\,C\,R\).

The modes \(L\) and \(R\) are the causally inaccessible partner modes
associated with the BEH and CEH, respectively.
For observers confined to the static region, the physically
accessible state is obtained by tracing over both inaccessible modes,
\begin{equation}
\rho_{ABC}
=
\operatorname{Tr}_{LR}
\left(
\rho_{ALBCR}
\right).
\label{eq:rho_ABC_reduced}
\end{equation}
The remaining three reduced states are obtained by tracing over the
corresponding complementary modes,
\begin{equation}
\begin{aligned}
\rho_{ABR}
&=
\operatorname{Tr}_{LC}
\left(
\rho_{ALBCR}
\right),
\\
\rho_{LBC}
&=
\operatorname{Tr}_{AR}
\left(
\rho_{ALBCR}
\right),
\\
\rho_{LBR}
&=
\operatorname{Tr}_{AC}
\left(
\rho_{ALBCR}
\right).
\end{aligned}
\label{eq:inaccessible_reduced_states}
\end{equation}
Accordingly, \(\rho_{ABC}\) is the physically accessible state,
whereas
\(\rho_{ABR}\), \(\rho_{LBC}\), and \(\rho_{LBR}\)
are physically inaccessible reduced states.
All four reduced density matrices have the three qubit X state form,
which allows their genuine multipartite entanglement and tripartite
nonlocality to be evaluated analytically.

\subsection{Genuine multipartite entanglement}
\label{sec:GME}

The four reduced states obtained in
Sec.~\ref{sec:state} possess the three qubit X state structure.
This allows their genuine multipartite entanglement to be evaluated
analytically by means of the genuine multipartite concurrence
\cite{Ma2011,Hashemi2012}. For a general three qubit X state written
in the computational basis,
\begin{equation}
\rho_X
=
\begin{pmatrix}
a_1 & 0 & 0 & 0 & 0 & 0 & 0 & z_1 \\
0 & a_2 & 0 & 0 & 0 & 0 & z_2 & 0 \\
0 & 0 & a_3 & 0 & 0 & z_3 & 0 & 0 \\
0 & 0 & 0 & a_4 & z_4 & 0 & 0 & 0 \\
0 & 0 & 0 & z_4^{*} & b_4 & 0 & 0 & 0 \\
0 & 0 & z_3^{*} & 0 & 0 & b_3 & 0 & 0 \\
0 & z_2^{*} & 0 & 0 & 0 & 0 & b_2 & 0 \\
z_1^{*} & 0 & 0 & 0 & 0 & 0 & 0 & b_1
\end{pmatrix},
\label{eq:three_qubit_X_state}
\end{equation}
where \(a_i\) and \(b_i\) denote the diagonal elements and \(z_i\)
the corresponding antidiagonal elements, with
\(i=1,2,3,4\). The density matrix conditions require
\begin{equation}
a_i\geq0,
\qquad
b_i\geq0,
\qquad
\sum_{i=1}^{4}(a_i+b_i)=1,
\qquad
|z_i|^{2}\leq a_i b_i.
\label{eq:X_state_conditions}
\end{equation}

For each antidiagonal pair, define
\begin{equation}
m_i
=
\sum_{j\neq i}
\sqrt{a_j b_j},
\qquad
i,j=1,2,3,4.
\label{eq:mi_definition}
\end{equation}
The genuine multipartite concurrence of a three qubit X state is
then given by
\cite{Hashemi2012}
\begin{equation}
C_{\mathrm{GM}}(\rho_X)
=
2\max
\left\{
0,\,
\max_{1\leq i\leq4}
\left(
|z_i|-m_i
\right)
\right\}.
\label{eq:general_GM_concurrence}
\end{equation}
Accordingly, \(C_{\mathrm{GM}}>0\) certifies genuine multipartite
entanglement.

The explicit density matrices of
\(\rho_{ABC}\), \(\rho_{ABR}\), \(\rho_{LBC}\), and
\(\rho_{LBR}\) are given in
Appendix~\ref{app:matrices}.
Each of these states contains a single nonzero antidiagonal pair.
Substitution of the corresponding matrix elements into
Eq.~(\ref{eq:general_GM_concurrence}) therefore yields the following
closed expressions.

\begin{widetext}

\begin{equation}
\resizebox{0.96\textwidth}{!}{$\displaystyle
C_{\mathrm{GM}}(\rho_{ABC})
=
2\mu_{\mathcal{H}}\mu_{\mathcal{C}}
\max\left\{
0,\,
p\alpha\beta
-
\sqrt{
q(1+\nu_{\mathcal{H}}^{2})
\left[
p\alpha^{2}\nu_{\mathcal{C}}^{2}
+
q(1+\nu_{\mathcal{C}}^{2})
\right]
}
-
\sqrt{
q
\left[
p\alpha^{2}\nu_{\mathcal{H}}^{2}\nu_{\mathcal{C}}^{2}
+
q(1+\nu_{\mathcal{H}}^{2})(1+\nu_{\mathcal{C}}^{2})
\right]
}
-
\sqrt{
q(1+\nu_{\mathcal{C}}^{2})
\left[
p\alpha^{2}\nu_{\mathcal{H}}^{2}
+
q(1+\nu_{\mathcal{H}}^{2})
\right]
}
\right\}.
$}
\label{eq:IV_CGM_ABC}
\end{equation}

\begin{equation}
\resizebox{0.96\textwidth}{!}{$\displaystyle
C_{\mathrm{GM}}(\rho_{ABR})
=
2\mu_{\mathcal{H}}\nu_{\mathcal{C}}
\max\left\{
0,\,
p\alpha\beta
-
\sqrt{
q(1+\nu_{\mathcal{H}}^{2})
\left[
p\alpha^{2}\mu_{\mathcal{C}}^{2}
+
q(1+\mu_{\mathcal{C}}^{2})
\right]
}
-
\sqrt{
q(1+\mu_{\mathcal{C}}^{2})
\left[
p\alpha^{2}\nu_{\mathcal{H}}^{2}
+
q(1+\nu_{\mathcal{H}}^{2})
\right]
}
-
\sqrt{
q
\left[
p\alpha^{2}\nu_{\mathcal{H}}^{2}\mu_{\mathcal{C}}^{2}
+
q(1+\nu_{\mathcal{H}}^{2})(1+\mu_{\mathcal{C}}^{2})
\right]
}
\right\}.
$}
\label{eq:IV_CGM_ABR}
\end{equation}

\begin{equation}
\resizebox{0.96\textwidth}{!}{$\displaystyle
C_{\mathrm{GM}}(\rho_{LBC})
=
2\nu_{\mathcal{H}}\mu_{\mathcal{C}}
\max\left\{
0,\,
p\alpha\beta
-
\sqrt{
q(1+\nu_{\mathcal{C}}^{2})
\left[
p\alpha^{2}\mu_{\mathcal{H}}^{2}
+
q(1+\mu_{\mathcal{H}}^{2})
\right]
}
-
\sqrt{
q
\left[
p\alpha^{2}\mu_{\mathcal{H}}^{2}\nu_{\mathcal{C}}^{2}
+
q(1+\mu_{\mathcal{H}}^{2})(1+\nu_{\mathcal{C}}^{2})
\right]
}
-
\sqrt{
q(1+\mu_{\mathcal{H}}^{2})
\left[
p\alpha^{2}\nu_{\mathcal{C}}^{2}
+
q(1+\nu_{\mathcal{C}}^{2})
\right]
}
\right\}.
$}
\label{eq:IV_CGM_LBC}
\end{equation}

\begin{equation}
\resizebox{0.96\textwidth}{!}{$\displaystyle
C_{\mathrm{GM}}(\rho_{LBR})
=
2\nu_{\mathcal{H}}\nu_{\mathcal{C}}
\max\left\{
0,\,
p\alpha\beta
-
\sqrt{
q
\left[
p\alpha^{2}\mu_{\mathcal{H}}^{2}\mu_{\mathcal{C}}^{2}
+
q(1+\mu_{\mathcal{H}}^{2})(1+\mu_{\mathcal{C}}^{2})
\right]
}
-
\sqrt{
q(1+\mu_{\mathcal{C}}^{2})
\left[
p\alpha^{2}\mu_{\mathcal{H}}^{2}
+
q(1+\mu_{\mathcal{H}}^{2})
\right]
}
-
\sqrt{
q(1+\mu_{\mathcal{H}}^{2})
\left[
p\alpha^{2}\mu_{\mathcal{C}}^{2}
+
q(1+\mu_{\mathcal{C}}^{2})
\right]
}
\right\}.
$}
\label{eq:IV_CGM_LBR}
\end{equation}

\end{widetext}

Eqs.~(\ref{eq:IV_CGM_ABC})--(\ref{eq:IV_CGM_LBR}) show explicitly
that the genuine multipartite concurrence depends on both horizon
transformations as well as on the initial-state parameters
\(p\) and \(\alpha\).
The role of the two horizons becomes particularly transparent in
the pure-state limit.
For \(p=1\), one has \(q=0\), and all white-noise contributions
vanish. The four concurrences then reduce to
\begin{equation}
\begin{aligned}
C_{\mathrm{GM}}(\rho_{ABC})
&=
2\alpha\beta
\mu_{\mathcal{H}}\mu_{\mathcal{C}},
\\
C_{\mathrm{GM}}(\rho_{ABR})
&=
2\alpha\beta
\mu_{\mathcal{H}}\nu_{\mathcal{C}},
\\
C_{\mathrm{GM}}(\rho_{LBC})
&=
2\alpha\beta
\nu_{\mathcal{H}}\mu_{\mathcal{C}},
\\
C_{\mathrm{GM}}(\rho_{LBR})
&=
2\alpha\beta
\nu_{\mathcal{H}}\nu_{\mathcal{C}}.
\end{aligned}
\label{eq:IV_pure_CGM}
\end{equation}

Within the surface-gravity normalization adopted here, the Nariai
limit \(x\rightarrow1\) gives
\(\mu_{\mathcal H},\mu_{\mathcal C}\rightarrow1\) and
\(\nu_{\mathcal H},\nu_{\mathcal C}\rightarrow0\).
Eq.~(\ref{eq:IV_pure_CGM}) then shows that, for a pure initial
state,
\(C_{\mathrm{GM}}(\rho_{ABC})\rightarrow2\alpha\beta\),
whereas
\(C_{\mathrm{GM}}(\rho_{ABR})\),
\(C_{\mathrm{GM}}(\rho_{LBC})\), and
\(C_{\mathrm{GM}}(\rho_{LBR})\) vanish.
Thus, in this limit, the genuine multipartite concurrence becomes
concentrated in the physically accessible sector.


\subsection{Genuine tripartite nonlocality}
\label{sec:nonlocality}

Having characterized the genuine multipartite entanglement of the
four reduced states, it is natural to ask whether the surviving
correlations also exhibit genuine tripartite nonlocality.
Entanglement alone does not guarantee such nonlocality.
For three parties, genuine tripartite nonlocality can be certified
through violation of the Svetlichny inequality
\cite{Svetlichny1987,AjoyRungta2010,
Wang2020GenuineNonlocality}.

For two dichotomic observables at each site, the Svetlichny operator
can be written as
\begin{align}
\widehat{\mathcal{S}}
={}&
A_0(B_0+B_1)C_0
+
A_0(B_0-B_1)C_1
\notag\\
&+
A_1(B_0-B_1)C_0
-
A_1(B_0+B_1)C_1,
\label{eq:Svetlichny_operator}
\end{align}
where \(A_i\), \(B_i\), and \(C_i\) denote local dichotomic
observables of the three parties.
Any state admitting a hybrid local and nonlocal hiddenvariable 
description satisfies
\begin{equation}
\left|
\Tr
\left(
\rho\widehat{\mathcal{S}}
\right)
\right|
\leq4.
\label{eq:Svetlichny_bound}
\end{equation}
The optimized Svetlichny parameter is therefore defined as
\begin{equation}
\mathcal{S}_{\max}(\rho)
=
\max_{\mathcal{M}}
\left|
\Tr
\left(
\rho\widehat{\mathcal{S}}
\right)
\right|,
\label{eq:Smax_definition}
\end{equation}
where \(\mathcal{M}\) denotes optimization over all local measurement
directions. A value
\(\mathcal{S}_{\max}>4\) certifies genuine tripartite nonlocality
within the Svetlichny criterion.

For the single antidiagonal coherence three qubit X states considered
here, the optimization can be carried out analytically
\cite{Wang2020GenuineNonlocality}, giving
\begin{equation}
\mathcal{S}_{\max}(\rho_X)
=
\max
\left\{
8\sqrt{2}
\max_{1\leq i\leq4}|z_i|,
\,
4|\mathcal{N}|
\right\},
\label{eq:Smax_general_X}
\end{equation}
with
\begin{equation}
\mathcal{N}
=
\Tr
\left[
\rho_X
\left(
\sigma_z\otimes\sigma_z\otimes\sigma_z
\right)
\right].
\label{eq:N_definition}
\end{equation}
Since
\(\sigma_z\otimes\sigma_z\otimes\sigma_z\)
has eigenvalues \(\pm1\), one has
\(|\mathcal{N}|\leq1\), and hence
\(4|\mathcal{N}|\leq4\).
The second branch of
Eq.~(\ref{eq:Smax_general_X}) can therefore never by itself produce
a Svetlichny violation.
Consequently, whether a violation occurs is determined entirely by
whether the first branch exceeds \(4\).

To obtain the optimized Svetlichny parameters of the four reduced
states, it is convenient to introduce
\begin{equation}
\Delta_{\mathcal H}
=
\mu_{\mathcal H}^{2}
-
\nu_{\mathcal H}^{2},
\qquad
\Delta_{\mathcal C}
=
\mu_{\mathcal C}^{2}
-
\nu_{\mathcal C}^{2}.
\label{eq:Delta_HC}
\end{equation}

The matrix elements of the four reduced states
\(\rho_{ABC}\), \(\rho_{ABR}\), \(\rho_{LBC}\), and
\(\rho_{LBR}\) are given explicitly in
Appendix~\ref{app:matrices}.
Substituting these matrix elements into
Eq.~(\ref{eq:Smax_general_X}) yields the optimized Svetlichny
parameters for the four reduced states.
\begin{equation}
\begin{aligned}
\mathcal{S}_{\max}(\rho_{ABC})
=
\max\Bigl\{
&
8\sqrt{2}\,
p\alpha\beta
\mu_{\mathcal H}\mu_{\mathcal C},
\\
&
4p
\left|
\alpha^{2}
\Delta_{\mathcal H}
\Delta_{\mathcal C}
-
\beta^{2}
\right|
\Bigr\}.
\end{aligned}
\label{eq:IV_S_ABC}
\end{equation}

For the three physically inaccessible reduced states
\(\rho_{ABR}\), \(\rho_{LBC}\), and \(\rho_{LBR}\), the optimized
Svetlichny parameters are given by
\begin{equation}
\begin{aligned}
\mathcal{S}_{\max}(\rho_{ABR})
=
\max\Bigl\{
&
8\sqrt{2}\,
p\alpha\beta
\mu_{\mathcal H}\nu_{\mathcal C},
\\
&
4p
\left|
\alpha^{2}
\Delta_{\mathcal H}
\Delta_{\mathcal C}
+
\beta^{2}
\right|
\Bigr\},
\end{aligned}
\label{eq:IV_S_ABR}
\end{equation}

\begin{equation}
\begin{aligned}
\mathcal{S}_{\max}(\rho_{LBC})
=
\max\Bigl\{
&
8\sqrt{2}\,
p\alpha\beta
\nu_{\mathcal H}\mu_{\mathcal C},
\\
&
4p
\left|
\alpha^{2}
\Delta_{\mathcal H}
\Delta_{\mathcal C}
+
\beta^{2}
\right|
\Bigr\},
\end{aligned}
\label{eq:IV_S_LBC}
\end{equation}

and
\begin{equation}
\begin{aligned}
\mathcal{S}_{\max}(\rho_{LBR})
=
\max\Bigl\{
&
8\sqrt{2}\,
p\alpha\beta
\nu_{\mathcal H}\nu_{\mathcal C},
\\
&
4p
\left|
\alpha^{2}
\Delta_{\mathcal H}
\Delta_{\mathcal C}
-
\beta^{2}
\right|
\Bigr\}.
\end{aligned}
\label{eq:IV_S_LBR}
\end{equation}

These expressions allow the possibility of Svetlichny violation to
be determined analytically before any numerical analysis is carried
out.
For fermionic modes,
\begin{equation}
\frac{1}{\sqrt{2}}
\leq
\mu_{\mathcal H},
\mu_{\mathcal C}
\leq1,
\qquad
0
\leq
\nu_{\mathcal H},
\nu_{\mathcal C}
\leq
\frac{1}{\sqrt{2}},
\label{eq:fermionic_bounds}
\end{equation}
while
\(0\leq p\leq1\) and
\(\alpha\beta\leq1/2\).
It follows that the first branches of the three physically
inaccessible states satisfy
\begin{equation}
\begin{aligned}
8\sqrt{2}\,
p\alpha\beta
\mu_{\mathcal H}\nu_{\mathcal C}
&\leq4,
\\
8\sqrt{2}\,
p\alpha\beta
\nu_{\mathcal H}\mu_{\mathcal C}
&\leq4,
\\
8\sqrt{2}\,
p\alpha\beta
\nu_{\mathcal H}\nu_{\mathcal C}
&\leq2\sqrt{2}.
\end{aligned}
\label{eq:IV_inaccessible_antidiagonal_bounds}
\end{equation}
Together with the bound on the second branch, these inequalities
imply
\begin{equation}
\mathcal{S}_{\max}(\rho_X)
\leq4,
\qquad
\rho_X
\in
\left\{
\rho_{ABR},
\rho_{LBC},
\rho_{LBR}
\right\}.
\label{eq:IV_inaccessible_S_bound}
\end{equation}
Thus, within the fermionic parameter range considered here, none of
the three physically inaccessible reduced states can violate the
Svetlichny inequality, even though their genuine multipartite
concurrence can remain nonzero.

The physically accessible state behaves differently.
Because the second branch of
Eq.~(\ref{eq:IV_S_ABC}) is bounded by \(4\), a Svetlichny violation
occurs if and only if
\begin{equation}
p\alpha\beta
\mu_{\mathcal H}\mu_{\mathcal C}
>
\frac{1}{2\sqrt{2}}.
\label{eq:IV_nonlocal_condition}
\end{equation}
Whenever this condition is satisfied,
\begin{equation}
\mathcal{S}_{\max}(\rho_{ABC})
=
8\sqrt{2}\,
p\alpha\beta
\mu_{\mathcal H}\mu_{\mathcal C}
>
4,
\label{eq:IV_S_ABC_violation}
\end{equation}
so that genuine tripartite nonlocality is certified for the
physically accessible state.

The distinction between genuine multipartite entanglement and
Svetlichny nonlocality becomes particularly transparent for the pure
maximally coherent GHZ case,
\(p=1\) and
\(\alpha=\beta=1/\sqrt{2}\).
Eq.~(\ref{eq:IV_pure_CGM}) then gives
\begin{equation}
C_{\mathrm{GM}}(\rho_{ABC})
=
\mu_{\mathcal H}\mu_{\mathcal C},
\label{eq:IV_CGM_ABC_pure}
\end{equation}
whereas
Eq.~(\ref{eq:IV_S_ABC}) yields
\begin{equation}
\mathcal{S}_{\max}(\rho_{ABC})
=
4\sqrt{2}\,
\mu_{\mathcal H}\mu_{\mathcal C}.
\label{eq:IV_S_ABC_pure}
\end{equation}
Hence,
\begin{equation}
\mathcal{S}_{\max}(\rho_{ABC})
=
4\sqrt{2}\,
C_{\mathrm{GM}}(\rho_{ABC}).
\label{eq:IV_S_C_relation}
\end{equation}
Although any positive
\(C_{\mathrm{GM}}(\rho_{ABC})\) indicates genuine multipartite
entanglement, Svetlichny violation requires the stronger condition
\begin{equation}
C_{\mathrm{GM}}(\rho_{ABC})
>
\frac{1}{\sqrt{2}}.
\label{eq:IV_CGM_S_threshold}
\end{equation}
There is therefore a finite parameter regime in which the physically
accessible state remains genuinely multipartite entangled but does
not exhibit a Svetlichny violation.
This analytical distinction provides the basis for the direct
comparison between genuine multipartite entanglement and genuine
tripartite nonlocality presented in
Sec.~\ref{sec:results}.


\section{Numerical results and discussion}
\label{sec:results}

For the numerical analysis, the fermionic modes associated with the
black-hole event horizon (BEH) and the cosmological event horizon
(CEH) are taken to have the same frequency,
\(\omega_A=\omega_C=\omega\).
It is convenient to introduce the dimensionless spacetime parameter
\(x=3M\sqrt{\Lambda}\) and the dimensionless field frequency
\(\varpi=\omega/\sqrt{\Lambda}\).
The physical Schwarzschild--de Sitter region is restricted to
\(0<x<1\), while the limit \(x\rightarrow1\) corresponds to the
Nariai limit.


\subsection{Genuine multipartite concurrence}
\label{sec:results_GME}

\begin{figure}[!t]
\centering
\includegraphics[
width=0.98\columnwidth
]{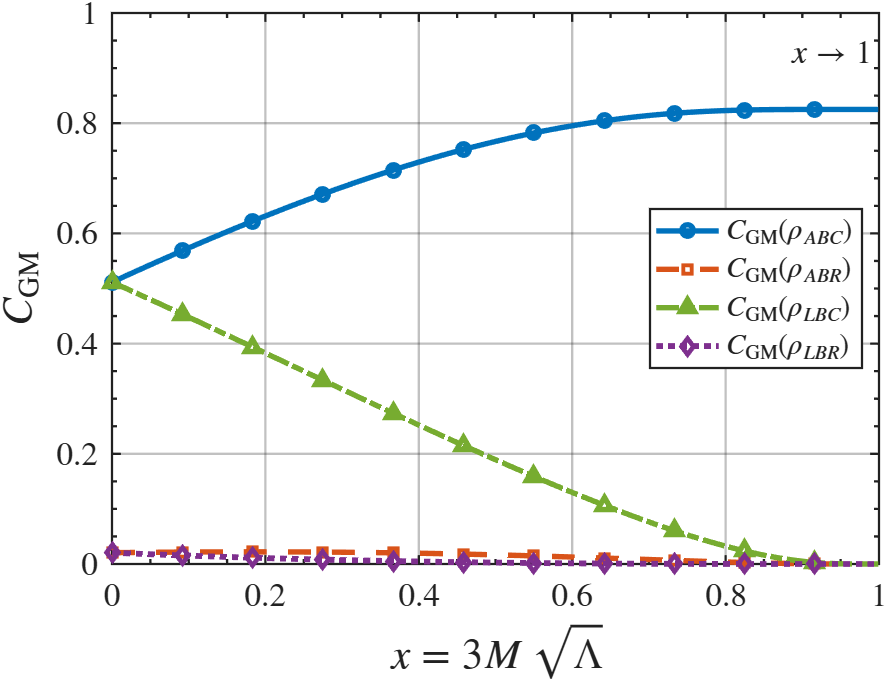}
\caption{
Genuine multipartite concurrence \(C_{\mathrm{GM}}\) of the
physically accessible state \(\rho_{ABC}\) and the physically
inaccessible reduced states
\(\rho_{ABR}\), \(\rho_{LBC}\), and \(\rho_{LBR}\)
as a function of \(x=3M\sqrt{\Lambda}\), for
\(p=0.9\), \(\alpha=\beta=1/\sqrt{2}\), and \(\varpi=0.5\).
}
\label{fig:CGM_x}
\end{figure}

We first examine how genuine multipartite entanglement is distributed
among the physically accessible and physically inaccessible reduced
states as the spacetime parameter varies.
Fig.~\ref{fig:CGM_x} displays the genuine multipartite concurrence
of the physically accessible state \(\rho_{ABC}\) and the three
physically inaccessible reduced states
\(\rho_{ABR}\), \(\rho_{LBC}\), and \(\rho_{LBR}\)
as functions of \(x=3M\sqrt{\Lambda}\).
The parameters are fixed at
\(p=0.9\), \(\alpha=\beta=1/\sqrt{2}\), and \(\varpi=0.5\).

As shown in Fig.~\ref{fig:CGM_x},
\(C_{\mathrm{GM}}(\rho_{ABC})\) increases monotonically with \(x\)
and gradually approaches a finite plateau.
By contrast,
\(C_{\mathrm{GM}}(\rho_{LBC})\) decreases continuously and tends
to zero as \(x\) approaches the Nariai limit.
As \(x\rightarrow0\),
\(C_{\mathrm{GM}}(\rho_{ABC})\) and
\(C_{\mathrm{GM}}(\rho_{LBC})\) approach the same value.
The two curves then separate and evolve in opposite directions as
\(x\) increases.

A similar coincidence is observed for
\(C_{\mathrm{GM}}(\rho_{ABR})\) and
\(C_{\mathrm{GM}}(\rho_{LBR})\) as \(x\rightarrow0\).
The pairwise coincidence of
\(\rho_{ABC}\) with \(\rho_{LBC}\), and of
\(\rho_{ABR}\) with \(\rho_{LBR}\), in this limit is demonstrated
analytically in Appendix~\ref{app:common_origin}.

With increasing \(x\),
\(C_{\mathrm{GM}}(\rho_{ABR})\) and
\(C_{\mathrm{GM}}(\rho_{LBR})\) decrease and become increasingly
small.
The concurrence of \(\rho_{LBC}\) also decreases continuously.
As \(x\) approaches the Nariai limit, the genuine multipartite
concurrences of all three physically inaccessible reduced states
tend to zero, whereas
\(C_{\mathrm{GM}}(\rho_{ABC})\) remains finite.

\begin{figure*}[!t]
\centering
\includegraphics[
width=0.98\textwidth
]{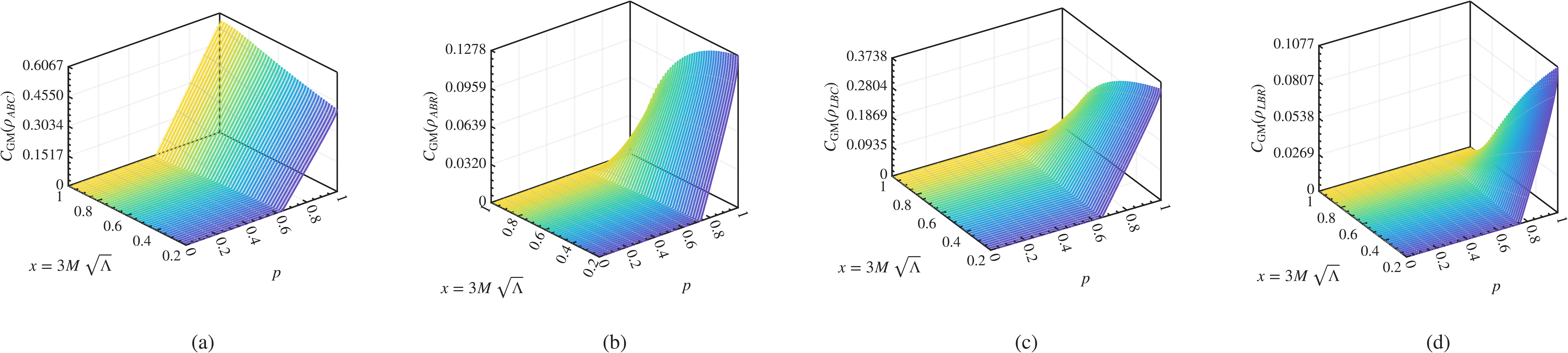}
\caption{
Genuine multipartite concurrence \(C_{\mathrm{GM}}\) as functions of
the mixing parameter \(p\) and the spacetime parameter
\(x=3M\sqrt{\Lambda}\), for
\(\alpha=\beta=1/\sqrt{2}\) and \(\varpi=0.5\).
Panels (a)--(d) correspond to
\(\rho_{ABC}\), \(\rho_{ABR}\), \(\rho_{LBC}\), and
\(\rho_{LBR}\), respectively.
}
\label{fig:CGM_p_x}
\end{figure*}

The dependence on the initial white noise contribution is examined
by varying the mixing parameter \(p\) together with the spacetime
parameter \(x\).
Fig.~\ref{fig:CGM_p_x} shows the resulting concurrence surfaces
for the four reduced states, with
\(\alpha=\beta=1/\sqrt{2}\) and \(\varpi=0.5\).

Fig.~\ref{fig:CGM_p_x} shows that the three physically inaccessible
reduced states
\(\rho_{ABR}\), \(\rho_{LBC}\), and \(\rho_{LBR}\)
display a similar overall dependence on \(p\) and \(x\).
For a fixed value of \(p\), their genuine multipartite concurrences
decrease as \(x\) increases and become small in the large \(x\)
region.
For a fixed value of \(x\), the concurrence increases with \(p\)
once it becomes nonzero.
The variation with \(p\) is more evident in the small \(x\) region
and becomes weaker as \(x\) increases.

The physically accessible state \(\rho_{ABC}\) displays a different
dependence on \(x\).
For fixed \(p\),
\(C_{\mathrm{GM}}(\rho_{ABC})\) increases with \(x\).
For fixed \(x\), it also increases as \(p\) increases.
The four panels therefore show a clear contrast between the
physically accessible state and the three physically inaccessible
reduced states as the spacetime parameter is varied.

\begin{figure*}[!t]
\centering
\includegraphics[
width=0.98\textwidth
]{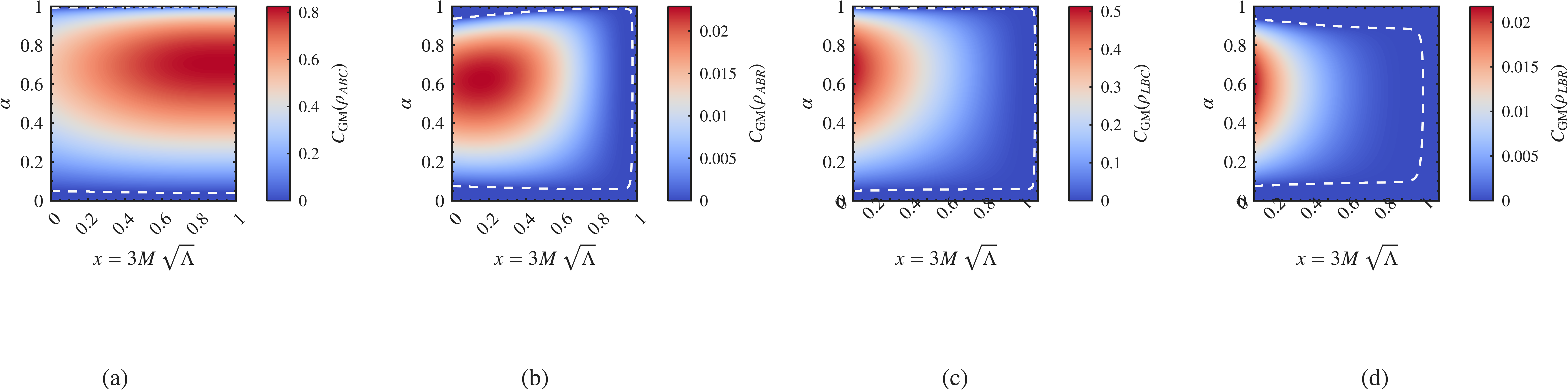}
\caption{
Genuine multipartite concurrence \(C_{\mathrm{GM}}\) as functions of
the spacetime parameter \(x=3M\sqrt{\Lambda}\) and the initial-state
parameter \(\alpha\), for
\(p=0.9\) and \(\varpi=0.5\).
Panels (a)--(d) correspond to
\(\rho_{ABC}\), \(\rho_{ABR}\), \(\rho_{LBC}\), and
\(\rho_{LBR}\), respectively.
The white dashed curves delimit the regions of nonzero genuine
multipartite concurrence.
}
\label{fig:CGM_alpha_x}
\end{figure*}

The dependence on the initial generalized GHZ parameter is examined
by varying \(\alpha\).
Fig.~\ref{fig:CGM_alpha_x} shows
\(C_{\mathrm{GM}}\) in the \(x\)-\(\alpha\) parameter plane for
\(p=0.9\) and \(\varpi=0.5\), with
\(\beta=\sqrt{1-\alpha^{2}}\).

As shown in Fig.~\ref{fig:CGM_alpha_x},
\(C_{\mathrm{GM}}(\rho_{ABC})\) generally increases with \(x\)
for intermediate values of \(\alpha\).
The concurrences of
\(\rho_{ABR}\), \(\rho_{LBC}\), and
\(\rho_{LBR}\) show the opposite overall variation and decrease
as \(x\) increases.

For all four reduced states, the concurrence is small when
\(\alpha\) is close to either \(0\) or \(1\), while larger values
occur mainly at intermediate \(\alpha\).
The regions of nonzero concurrence are therefore confined to a
finite range of the initial state parameter.
For the three physically inaccessible reduced states, these regions
become progressively narrower as \(x\) increases.
This reduction is most pronounced for
\(\rho_{LBR}\), whose nonzero concurrence is mainly concentrated in
the small \(x\) region.

\begin{figure*}[!t]
\centering
\includegraphics[
width=0.98\textwidth
]{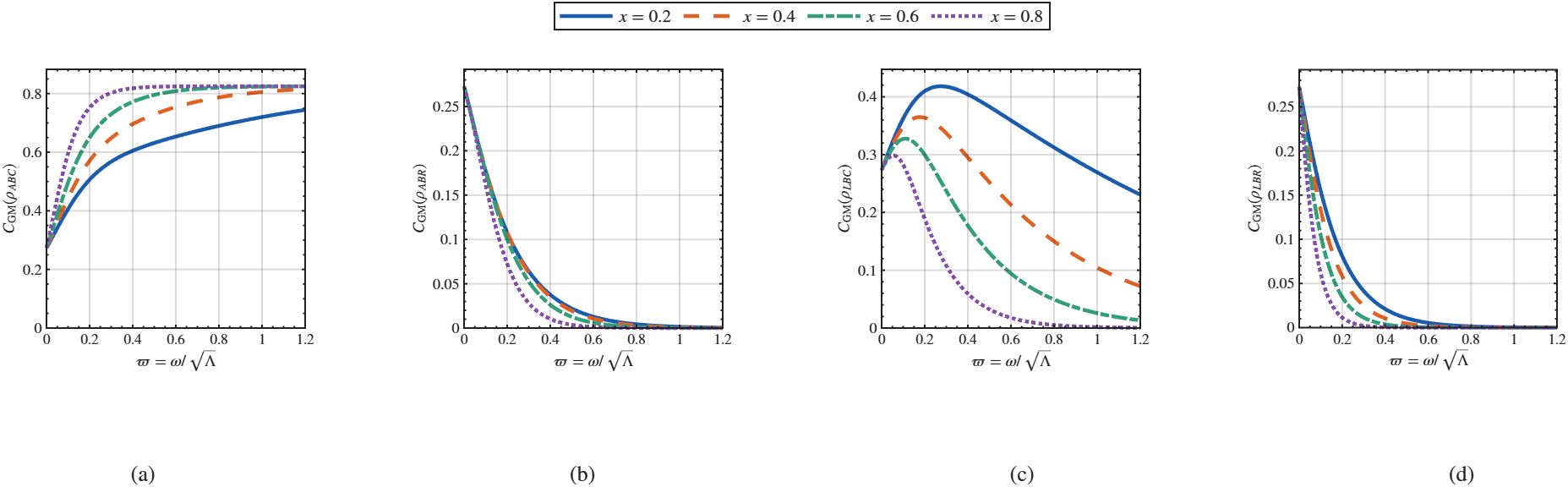}
\caption{
Genuine multipartite concurrence \(C_{\mathrm{GM}}\) as a function
of the dimensionless field frequency
\(\varpi=\omega/\sqrt{\Lambda}\), for
\(p=0.9\), \(\alpha=\beta=1/\sqrt{2}\), and
\(x=0.2,\,0.4,\,0.6,\) and \(0.8\).
Panels (a)--(d) correspond to
\(\rho_{ABC}\), \(\rho_{ABR}\), \(\rho_{LBC}\), and
\(\rho_{LBR}\), respectively.
}
\label{fig:CGM_varpi}
\end{figure*}

We next examine the dependence on the dimensionless field frequency.
Fig.~\ref{fig:CGM_varpi} shows the genuine multipartite concurrence
as a function of
\(\varpi=\omega/\sqrt{\Lambda}\).
The parameters are fixed at
\(p=0.9\) and \(\alpha=\beta=1/\sqrt{2}\), while
\(x=0.2,\,0.4,\,0.6,\) and \(0.8\) are considered.

As shown in Fig.~\ref{fig:CGM_varpi}, the curves corresponding to
different values of \(x\) start from the same value at
\(\varpi=0\) in each panel.
For the physically accessible state \(\rho_{ABC}\),
\(C_{\mathrm{GM}}(\rho_{ABC})\) increases monotonically with
\(\varpi\) and gradually approaches a plateau.
At a fixed nonzero frequency, a larger value of \(x\) generally
corresponds to a larger concurrence.

For the physically inaccessible reduced state \(\rho_{ABR}\),
\(C_{\mathrm{GM}}(\rho_{ABR})\) decreases monotonically with
\(\varpi\) and approaches zero at large frequencies.
The decrease becomes more rapid as \(x\) increases.

A different dependence is observed for \(\rho_{LBC}\).
Its concurrence first increases with \(\varpi\), reaches a maximum,
and then decreases.
The position of the maximum shifts toward lower frequencies as
\(x\) increases.
The value and position of the maximum therefore vary with the
spacetime parameter.

For \(\rho_{LBR}\),
\(C_{\mathrm{GM}}(\rho_{LBR})\) decreases monotonically with
\(\varpi\) and approaches zero within the plotted frequency range.
Larger values of \(x\) correspond to a faster decrease.

Figs.~\ref{fig:CGM_x}--\ref{fig:CGM_varpi} show that the four reduced
states display different dependences on the spacetime parameter,
the initial state parameters, and the field frequency.
The comparison with genuine tripartite nonlocality is considered
next.


\subsection{Genuine tripartite nonlocality}
\label{sec:results_nonlocality}

\begin{figure}[!t]
\centering
\includegraphics[
width=0.98\columnwidth
]{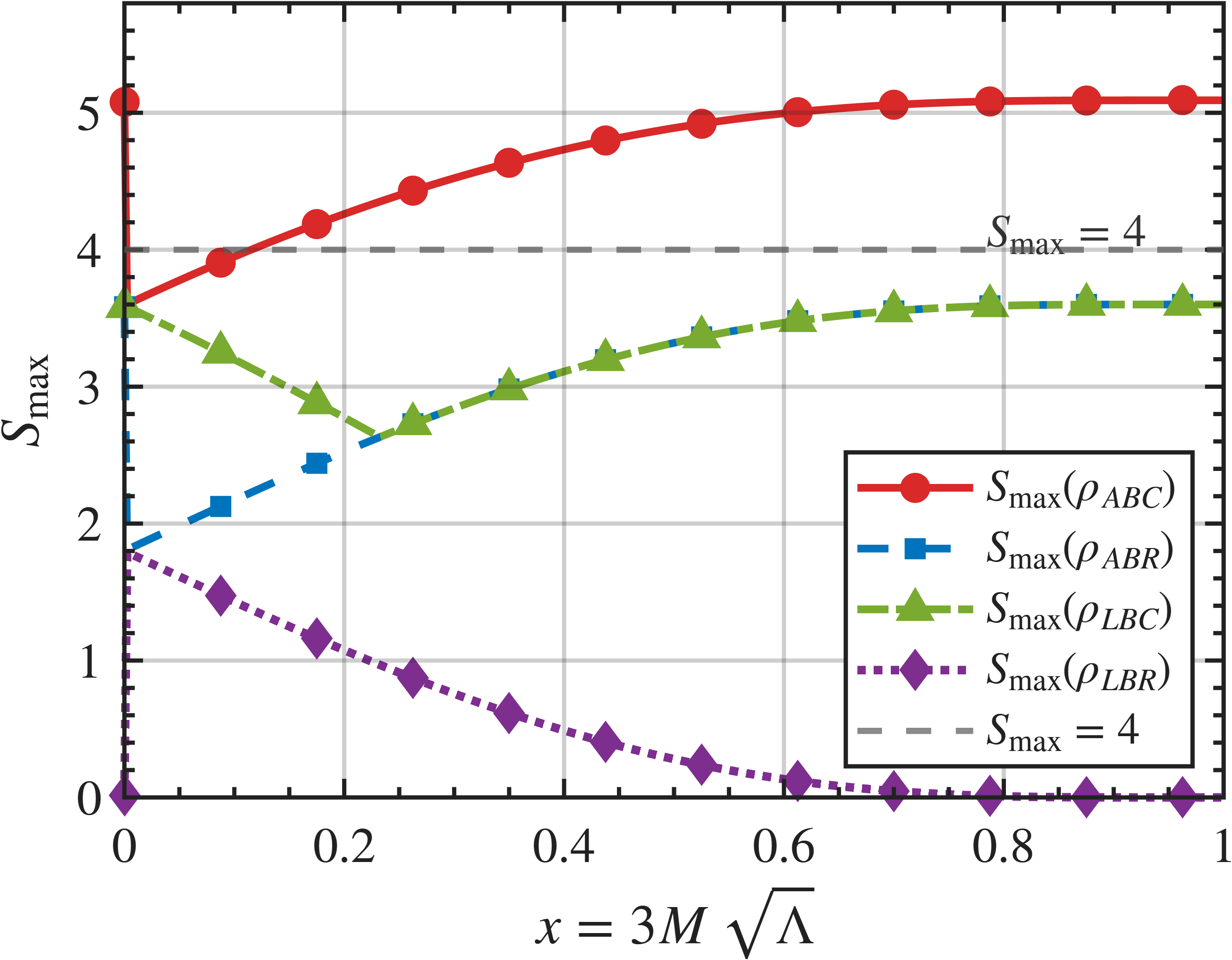}
\caption{
Optimized Svetlichny parameter \(S_{\max}\) of the
physically accessible state \(\rho_{ABC}\) and the physically
inaccessible reduced states
\(\rho_{ABR}\), \(\rho_{LBC}\), and \(\rho_{LBR}\)
as a function of \(x=3M\sqrt{\Lambda}\), for
\(p=0.9\), \(\alpha=\beta=1/\sqrt{2}\), and \(\varpi=0.5\).
The horizontal dashed line denotes the Svetlichny bound
\(S_{\max}=4\).
}
\label{fig:Smax_x}
\end{figure}

We now examine the optimized Svetlichny parameter and compare its
behavior with the genuine multipartite concurrence discussed above.
Fig.~\ref{fig:Smax_x} shows
\(S_{\max}\) for the four reduced states as a function of
\(x=3M\sqrt{\Lambda}\), with
\(p=0.9\), \(\alpha=\beta=1/\sqrt{2}\), and \(\varpi=0.5\).
The horizontal dashed line marks the Svetlichny bound
\(S_{\max}=4\).

As shown in Fig.~\ref{fig:Smax_x},
\(S_{\max}(\rho_{ABC})\) increases with \(x\) and crosses the
Svetlichny bound in the small \(x\) region.
It then increases more slowly and gradually approaches a finite
plateau.

The optimized Svetlichny parameters of the three physically
inaccessible reduced states remain below the Svetlichny bound over
the entire plotted interval.
For \(\rho_{LBC}\),
\(S_{\max}\) first decreases, reaches a minimum, and then increases
toward a plateau below \(4\).
For \(\rho_{ABR}\),
\(S_{\max}\) increases in the small-\(x\) region and subsequently
approaches the \(\rho_{LBC}\) curve.
The two curves overlap over the later part of the plotted interval.
For \(\rho_{LBR}\),
\(S_{\max}\) decreases monotonically with \(x\) and approaches zero.

The comparison with Fig.~\ref{fig:CGM_x} shows a clear difference
between genuine multipartite entanglement and genuine tripartite
nonlocality.
The three physically inaccessible reduced states can possess nonzero
genuine multipartite concurrence over part of the parameter range,
while their optimized Svetlichny parameters remain below the
Svetlichny bound.
Within the parameter range shown in Fig.~\ref{fig:Smax_x},
Svetlichny violation occurs only for the physically accessible state
\(\rho_{ABC}\).


\subsection{Joint influence of white noise and spacetime geometry}
\label{sec:results_joint}

\begin{figure}[!t]
\centering
\includegraphics[
width=0.98\columnwidth
]{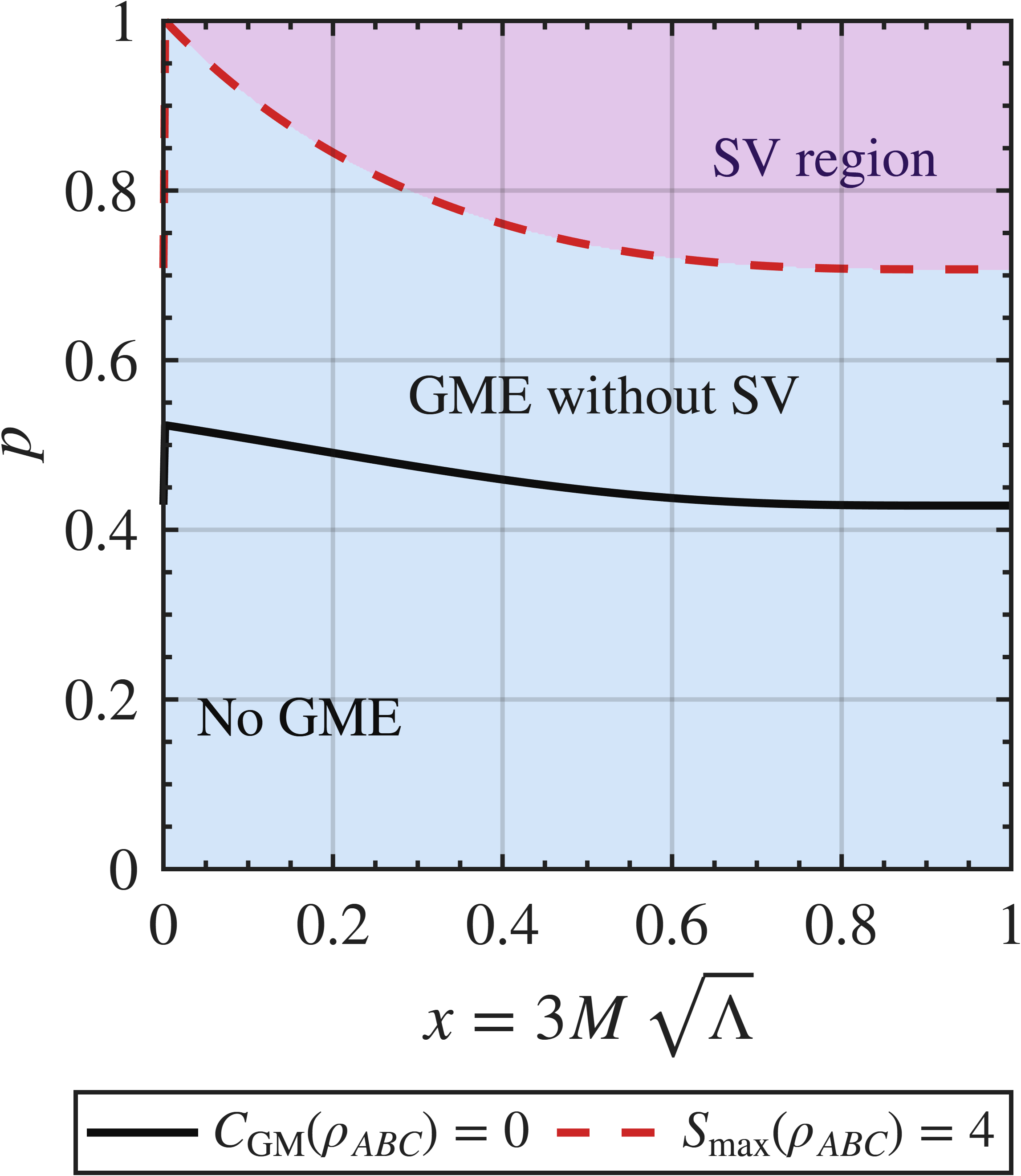}
\caption{
Parameter regions for the physically accessible state
\(\rho_{ABC}\) in the \(x\)-\(p\) plane, with
\(\alpha=\beta=1/\sqrt{2}\) and \(\varpi=0.5\).
The solid black curve denotes
\(C_{\mathrm{GM}}(\rho_{ABC})=0\), and the red dashed curve denotes
the Svetlichny threshold
\(S_{\max}(\rho_{ABC})=4\).
The three shaded regions correspond to the absence of genuine
multipartite entanglement, genuine multipartite entanglement without
Svetlichny violation, and Svetlichny violation, respectively.
}
\label{fig:GME_SV_region}
\end{figure}

The difference between the parameter regions supporting genuine
multipartite entanglement and Svetlichny violation can be seen more
directly in the \(x\)-\(p\) plane.
Fig.~\ref{fig:GME_SV_region} shows the corresponding regions for the
physically accessible state \(\rho_{ABC}\), with
\(\alpha=\beta=1/\sqrt{2}\) and \(\varpi=0.5\).
The solid black curve marks
\(C_{\mathrm{GM}}(\rho_{ABC})=0\), while the red dashed curve marks
the Svetlichny threshold
\(S_{\max}(\rho_{ABC})=4\).

As shown in Fig.~\ref{fig:GME_SV_region}, the boundary
\(C_{\mathrm{GM}}(\rho_{ABC})=0\) shifts toward smaller values of
\(p\) as \(x\) increases.
The minimum value of \(p\) required for nonzero genuine multipartite
concurrence therefore decreases gradually with \(x\).

The Svetlichny threshold
\(S_{\max}(\rho_{ABC})=4\) also shifts toward smaller values of
\(p\) as \(x\) increases.
Its variation with \(x\) is more pronounced than that of the
concurrence boundary.
The region above the red dashed curve corresponds to Svetlichny
violation, while the region below the solid black curve contains no
genuine multipartite entanglement.

The two boundaries remain separated over the plotted parameter
range.
Between them lies a finite region in which
\(C_{\mathrm{GM}}(\rho_{ABC})>0\) while
\(S_{\max}(\rho_{ABC})\leq4\).
The parameter region supporting genuine multipartite entanglement is
therefore broader than the region supporting Svetlichny violation.
This figure directly shows that genuine multipartite entanglement can
remain present even when the Svetlichny inequality is not violated.


\section{Conclusion}
\label{sec:conclusion}

Genuine multipartite entanglement and genuine tripartite
nonlocality have been investigated for a noisy generalized GHZ
state in Schwarzschild--de Sitter spacetime. The fermionic
transformations associated with the black hole event horizon and
the cosmological event horizon generate four three qubit reduced
states, consisting of the physically accessible state
\(\rho_{ABC}\) and the physically inaccessible reduced states
\(\rho_{ABR}\), \(\rho_{LBC}\), and \(\rho_{LBR}\).
Their genuine multipartite concurrence and optimized Svetlichny
parameter were obtained analytically and examined as functions of
the spacetime parameter, the initial-state parameters, and the
dimensionless field frequency.

The results show a clear difference between the physically
accessible and physically inaccessible sectors. As the spacetime
parameter \(x=3M\sqrt{\Lambda}\) increases,
\(C_{\mathrm{GM}}(\rho_{ABC})\) increases and approaches a finite
value, whereas the concurrences of
\(\rho_{ABR}\), \(\rho_{LBC}\), and \(\rho_{LBR}\) are suppressed
and vanish toward the large \(x\) region. Increasing the generalized
GHZ weight \(p\) generally enlarges the nonzero concurrence region,
while the dependence on the initial-state parameter \(\alpha\)
shows that appreciable genuine multipartite entanglement is mainly
concentrated at intermediate values of \(\alpha\).
For the frequency dependence considered here,
\(C_{\mathrm{GM}}(\rho_{ABC})\) increases with
\(\varpi\), whereas
\(C_{\mathrm{GM}}(\rho_{ABR})\) and
\(C_{\mathrm{GM}}(\rho_{LBR})\) decrease.
The concurrence of \(\rho_{LBC}\) instead exhibits a nonmonotonic
frequency dependence, increasing initially and decreasing after
reaching a maximum.

Genuine tripartite nonlocality displays a more restrictive
behavior. The optimized Svetlichny parameter of the physically
accessible state can exceed the Svetlichny bound, whereas those of
the three physically inaccessible reduced states remain at or below
the bound within the allowed fermionic parameter range considered
here. The \(x\)-\(p\) parameter diagram further shows a finite region
in which \(C_{\mathrm{GM}}(\rho_{ABC})\) is nonzero while the
Svetlichny inequality is not violated. Hence, genuine multipartite
entanglement can persist over a broader parameter region than
Svetlichny violation.

These results demonstrate that the two horizon transformations,
initial white noise, state preparation, and field frequency affect
genuine multipartite entanglement and genuine tripartite
nonlocality in distinct ways. In particular, multipartite
entanglement can remain present even when genuine tripartite
nonlocality is not certified, highlighting the different robustness
of the two quantum correlations in Schwarzschild--de Sitter
spacetime.


\begin{acknowledgments}
This work is supported by the Natural Science Foundation of Hainan
Province under Grant No.~125RC744 and the China Scholarship Council
(CSC).
\end{acknowledgments}

\appendix



\section{Explicit forms of the reduced density matrices}
\label{app:matrices}

For completeness, the explicit forms of the four three-qubit reduced
density matrices introduced in Sec.~\ref{sec:state} are presented
here. The computational basis is ordered as
\begin{equation}
\begin{aligned}
\mathcal{B}
=
\{
&\ket{000},
\ket{001},
\ket{010},
\ket{011},
\\
&\ket{100},
\ket{101},
\ket{110},
\ket{111}
\}.
\end{aligned}
\label{eq:app_basis}
\end{equation}
For convenience, we use
\(\beta=\sqrt{1-\alpha^{2}}\) and
\(q=(1-p)/8\).

\allowdisplaybreaks[2]


\subsection*{Physically accessible state \(\rho_{ABC}\)}

Tracing over the two causally inaccessible horizon-partner modes
\(L\) and \(R\) gives

\begingroup
\setlength{\arraycolsep}{2.2pt}
\renewcommand{\arraystretch}{1.05}
\begin{equation}
\resizebox{0.98\columnwidth}{!}{$\displaystyle
\rho_{ABC}
=
\begin{pmatrix}
d_{1}^{ABC} & 0 & 0 & 0 & 0 & 0 & 0 & \xi_{ABC} \\
0 & d_{2}^{ABC} & 0 & 0 & 0 & 0 & 0 & 0 \\
0 & 0 & d_{3}^{ABC} & 0 & 0 & 0 & 0 & 0 \\
0 & 0 & 0 & d_{4}^{ABC} & 0 & 0 & 0 & 0 \\
0 & 0 & 0 & 0 & d_{5}^{ABC} & 0 & 0 & 0 \\
0 & 0 & 0 & 0 & 0 & d_{6}^{ABC} & 0 & 0 \\
0 & 0 & 0 & 0 & 0 & 0 & d_{7}^{ABC} & 0 \\
\xi_{ABC} & 0 & 0 & 0 & 0 & 0 & 0 & d_{8}^{ABC}
\end{pmatrix}
$}
\label{eq:app_rho_ABC_matrix}
\end{equation}
\endgroup

The corresponding diagonal elements are
\begin{align}
d_{1}^{ABC}
&=
\mu_{\mathcal{H}}^{2}\mu_{\mathcal{C}}^{2}
\left(
p\alpha^{2}+q
\right),
\notag
\\[2pt]
d_{2}^{ABC}
&=
\mu_{\mathcal{H}}^{2}
\left[
p\alpha^{2}\nu_{\mathcal{C}}^{2}
+
q
\left(
1+\nu_{\mathcal{C}}^{2}
\right)
\right],
\notag
\\[2pt]
d_{3}^{ABC}
&=
q\mu_{\mathcal{H}}^{2}\mu_{\mathcal{C}}^{2},
\notag
\\[2pt]
d_{4}^{ABC}
&=
q\mu_{\mathcal{H}}^{2}
\left(
1+\nu_{\mathcal{C}}^{2}
\right),
\notag
\\[2pt]
d_{5}^{ABC}
&=
\mu_{\mathcal{C}}^{2}
\left[
p\alpha^{2}\nu_{\mathcal{H}}^{2}
+
q
\left(
1+\nu_{\mathcal{H}}^{2}
\right)
\right],
\notag
\\[2pt]
d_{6}^{ABC}
&=
p\alpha^{2}\nu_{\mathcal{H}}^{2}\nu_{\mathcal{C}}^{2}
\notag
\\
&\quad+
q
\left(
1+\nu_{\mathcal{H}}^{2}
\right)
\left(
1+\nu_{\mathcal{C}}^{2}
\right),
\notag
\\[2pt]
d_{7}^{ABC}
&=
q
\left(
1+\nu_{\mathcal{H}}^{2}
\right)
\mu_{\mathcal{C}}^{2},
\notag
\\[2pt]
d_{8}^{ABC}
&=
p\beta^{2}
+
q
\left(
1+\nu_{\mathcal{H}}^{2}
\right)
\left(
1+\nu_{\mathcal{C}}^{2}
\right).
\label{eq:app_d_ABC}
\end{align}

The only nonzero anti-diagonal pair is
\begin{equation}
\xi_{ABC}
=
p\alpha\beta
\mu_{\mathcal{H}}\mu_{\mathcal{C}}.
\label{eq:app_xi_ABC}
\end{equation}


\subsection*{Reduced state \(\rho_{ABR}\)}

Retaining the physically accessible BEH mode \(A\) and the
inaccessible CEH partner \(R\), while tracing over \(L\) and \(C\),
gives

\begingroup
\setlength{\arraycolsep}{2.2pt}
\renewcommand{\arraystretch}{1.05}
\begin{equation}
\resizebox{0.98\columnwidth}{!}{$\displaystyle
\rho_{ABR}
=
\begin{pmatrix}
d_{1}^{ABR} & 0 & 0 & 0 & 0 & 0 & 0 & 0 \\
0 & d_{2}^{ABR} & 0 & 0 & 0 & 0 & \xi_{ABR} & 0 \\
0 & 0 & d_{3}^{ABR} & 0 & 0 & 0 & 0 & 0 \\
0 & 0 & 0 & d_{4}^{ABR} & 0 & 0 & 0 & 0 \\
0 & 0 & 0 & 0 & d_{5}^{ABR} & 0 & 0 & 0 \\
0 & 0 & 0 & 0 & 0 & d_{6}^{ABR} & 0 & 0 \\
0 & \xi_{ABR} & 0 & 0 & 0 & 0 & d_{7}^{ABR} & 0 \\
0 & 0 & 0 & 0 & 0 & 0 & 0 & d_{8}^{ABR}
\end{pmatrix}
$}
\label{eq:app_rho_ABR_matrix}
\end{equation}
\endgroup

The corresponding diagonal elements are
\begin{align}
d_{1}^{ABR}
&=
\mu_{\mathcal{H}}^{2}
\left[
p\alpha^{2}\mu_{\mathcal{C}}^{2}
+
q
\left(
1+\mu_{\mathcal{C}}^{2}
\right)
\right],
\notag
\\[2pt]
d_{2}^{ABR}
&=
\mu_{\mathcal{H}}^{2}\nu_{\mathcal{C}}^{2}
\left(
p\alpha^{2}+q
\right),
\notag
\\[2pt]
d_{3}^{ABR}
&=
q\mu_{\mathcal{H}}^{2}
\left(
1+\mu_{\mathcal{C}}^{2}
\right),
\notag
\\[2pt]
d_{4}^{ABR}
&=
q\mu_{\mathcal{H}}^{2}\nu_{\mathcal{C}}^{2},
\notag
\\[2pt]
d_{5}^{ABR}
&=
p\alpha^{2}\nu_{\mathcal{H}}^{2}\mu_{\mathcal{C}}^{2}
\notag
\\
&\quad+
q
\left(
1+\nu_{\mathcal{H}}^{2}
\right)
\left(
1+\mu_{\mathcal{C}}^{2}
\right),
\notag
\\[2pt]
d_{6}^{ABR}
&=
\nu_{\mathcal{C}}^{2}
\left[
p\alpha^{2}\nu_{\mathcal{H}}^{2}
+
q
\left(
1+\nu_{\mathcal{H}}^{2}
\right)
\right],
\notag
\\[2pt]
d_{7}^{ABR}
&=
p\beta^{2}
+
q
\left(
1+\nu_{\mathcal{H}}^{2}
\right)
\left(
1+\mu_{\mathcal{C}}^{2}
\right),
\notag
\\[2pt]
d_{8}^{ABR}
&=
q
\left(
1+\nu_{\mathcal{H}}^{2}
\right)
\nu_{\mathcal{C}}^{2}.
\label{eq:app_d_ABR}
\end{align}

The only nonzero anti-diagonal pair is
\begin{equation}
\xi_{ABR}
=
p\alpha\beta
\mu_{\mathcal{H}}\nu_{\mathcal{C}}.
\label{eq:app_xi_ABR}
\end{equation}


\subsection*{Reduced state \(\rho_{LBC}\)}

Tracing over the physically accessible BEH mode \(A\) and the
inaccessible CEH partner \(R\) gives

\begingroup
\setlength{\arraycolsep}{2.2pt}
\renewcommand{\arraystretch}{1.05}
\begin{equation}
\resizebox{0.98\columnwidth}{!}{$\displaystyle
\rho_{LBC}
=
\begin{pmatrix}
d_{1}^{LBC} & 0 & 0 & 0 & 0 & 0 & 0 & 0 \\
0 & d_{2}^{LBC} & 0 & 0 & 0 & 0 & 0 & 0 \\
0 & 0 & d_{3}^{LBC} & 0 & 0 & 0 & 0 & 0 \\
0 & 0 & 0 & d_{4}^{LBC} & \xi_{LBC} & 0 & 0 & 0 \\
0 & 0 & 0 & \xi_{LBC} & d_{5}^{LBC} & 0 & 0 & 0 \\
0 & 0 & 0 & 0 & 0 & d_{6}^{LBC} & 0 & 0 \\
0 & 0 & 0 & 0 & 0 & 0 & d_{7}^{LBC} & 0 \\
0 & 0 & 0 & 0 & 0 & 0 & 0 & d_{8}^{LBC}
\end{pmatrix}
$}
\label{eq:app_rho_LBC_matrix}
\end{equation}
\endgroup

The corresponding diagonal elements are
\begin{align}
d_{1}^{LBC}
&=
\mu_{\mathcal{C}}^{2}
\left[
p\alpha^{2}\mu_{\mathcal{H}}^{2}
+
q
\left(
1+\mu_{\mathcal{H}}^{2}
\right)
\right],
\notag
\\[2pt]
d_{2}^{LBC}
&=
p\alpha^{2}\mu_{\mathcal{H}}^{2}\nu_{\mathcal{C}}^{2}
\notag
\\
&\quad+
q
\left(
1+\mu_{\mathcal{H}}^{2}
\right)
\left(
1+\nu_{\mathcal{C}}^{2}
\right),
\notag
\\[2pt]
d_{3}^{LBC}
&=
q
\left(
1+\mu_{\mathcal{H}}^{2}
\right)
\mu_{\mathcal{C}}^{2},
\notag
\\[2pt]
d_{4}^{LBC}
&=
p\beta^{2}
+
q
\left(
1+\mu_{\mathcal{H}}^{2}
\right)
\left(
1+\nu_{\mathcal{C}}^{2}
\right),
\notag
\\[2pt]
d_{5}^{LBC}
&=
\nu_{\mathcal{H}}^{2}\mu_{\mathcal{C}}^{2}
\left(
p\alpha^{2}+q
\right),
\notag
\\[2pt]
d_{6}^{LBC}
&=
\nu_{\mathcal{H}}^{2}
\left[
p\alpha^{2}\nu_{\mathcal{C}}^{2}
+
q
\left(
1+\nu_{\mathcal{C}}^{2}
\right)
\right],
\notag
\\[2pt]
d_{7}^{LBC}
&=
q\nu_{\mathcal{H}}^{2}\mu_{\mathcal{C}}^{2},
\notag
\\[2pt]
d_{8}^{LBC}
&=
q\nu_{\mathcal{H}}^{2}
\left(
1+\nu_{\mathcal{C}}^{2}
\right).
\label{eq:app_d_LBC}
\end{align}

The only nonzero anti-diagonal pair is
\begin{equation}
\xi_{LBC}
=
p\alpha\beta
\nu_{\mathcal{H}}\mu_{\mathcal{C}}.
\label{eq:app_xi_LBC}
\end{equation}


\subsection*{Fully inaccessible reduced state \(\rho_{LBR}\)}

Finally, tracing over both physically accessible horizon modes
\(A\) and \(C\) gives

\begingroup
\setlength{\arraycolsep}{2.2pt}
\renewcommand{\arraystretch}{1.05}
\begin{equation}
\resizebox{0.98\columnwidth}{!}{$\displaystyle
\rho_{LBR}
=
\begin{pmatrix}
d_{1}^{LBR} & 0 & 0 & 0 & 0 & 0 & 0 & 0 \\
0 & d_{2}^{LBR} & 0 & 0 & 0 & 0 & 0 & 0 \\
0 & 0 & d_{3}^{LBR} & 0 & 0 & \xi_{LBR} & 0 & 0 \\
0 & 0 & 0 & d_{4}^{LBR} & 0 & 0 & 0 & 0 \\
0 & 0 & 0 & 0 & d_{5}^{LBR} & 0 & 0 & 0 \\
0 & 0 & \xi_{LBR} & 0 & 0 & d_{6}^{LBR} & 0 & 0 \\
0 & 0 & 0 & 0 & 0 & 0 & d_{7}^{LBR} & 0 \\
0 & 0 & 0 & 0 & 0 & 0 & 0 & d_{8}^{LBR}
\end{pmatrix}
$}
\label{eq:app_rho_LBR_matrix}
\end{equation}
\endgroup

The corresponding diagonal elements are
\begin{align}
d_{1}^{LBR}
&=
p\alpha^{2}
\mu_{\mathcal{H}}^{2}
\mu_{\mathcal{C}}^{2}
\notag
\\
&\quad+
q
\left(
1+\mu_{\mathcal{H}}^{2}
\right)
\left(
1+\mu_{\mathcal{C}}^{2}
\right),
\notag
\\[2pt]
d_{2}^{LBR}
&=
\nu_{\mathcal{C}}^{2}
\left[
p\alpha^{2}\mu_{\mathcal{H}}^{2}
+
q
\left(
1+\mu_{\mathcal{H}}^{2}
\right)
\right],
\notag
\\[2pt]
d_{3}^{LBR}
&=
p\beta^{2}
+
q
\left(
1+\mu_{\mathcal{H}}^{2}
\right)
\left(
1+\mu_{\mathcal{C}}^{2}
\right),
\notag
\\[2pt]
d_{4}^{LBR}
&=
q
\left(
1+\mu_{\mathcal{H}}^{2}
\right)
\nu_{\mathcal{C}}^{2},
\notag
\\[2pt]
d_{5}^{LBR}
&=
\nu_{\mathcal{H}}^{2}
\left[
p\alpha^{2}\mu_{\mathcal{C}}^{2}
+
q
\left(
1+\mu_{\mathcal{C}}^{2}
\right)
\right],
\notag
\\[2pt]
d_{6}^{LBR}
&=
\nu_{\mathcal{H}}^{2}\nu_{\mathcal{C}}^{2}
\left(
p\alpha^{2}+q
\right),
\notag
\\[2pt]
d_{7}^{LBR}
&=
q\nu_{\mathcal{H}}^{2}
\left(
1+\mu_{\mathcal{C}}^{2}
\right),
\notag
\\[2pt]
d_{8}^{LBR}
&=
q
\nu_{\mathcal{H}}^{2}
\nu_{\mathcal{C}}^{2}.
\label{eq:app_d_LBR}
\end{align}

The only nonzero anti-diagonal pair is
\begin{equation}
\xi_{LBR}
=
p\alpha\beta
\nu_{\mathcal{H}}\nu_{\mathcal{C}}.
\label{eq:app_xi_LBR}
\end{equation}

Finally, the normalization of the four reduced density matrices
follows from the normalization of the two independent fermionic
transformations. Explicitly,
\begin{equation}
\begin{aligned}
\sum_{i=1}^{8} d_i^{XYZ}
&=1,
\\
XYZ
&\in
\left\{
ABC,\,ABR,\,LBC,\,LBR
\right\}.
\end{aligned}
\label{eq:app_normalization_check}
\end{equation}


\section{Pairwise coincidence of the genuine multipartite
concurrences in the \(x\rightarrow0\) limit}
\label{app:common_origin}

This appendix establishes the pairwise coincidence of the genuine
multipartite concurrences in the limit \(x\rightarrow0\).
Specifically, it is shown that
\(C_{\mathrm{GM}}(\rho_{ABC})\) and
\(C_{\mathrm{GM}}(\rho_{LBC})\) approach the same value, as do
\(C_{\mathrm{GM}}(\rho_{ABR})\) and
\(C_{\mathrm{GM}}(\rho_{LBR})\).
The result follows from the limiting behavior of the black-hole
horizon coefficients and does not rely on the particular choice
\(\alpha=\beta=1/\sqrt{2}\).

For
\[
x=3M\sqrt{\Lambda},
\]
the dimensionless black-hole and cosmological horizon radii are
\begin{equation}
\begin{aligned}
R_{\mathcal H}
&=
2\cos
\left[
\frac{\arccos(-x)+4\pi}{3}
\right],
\\[6pt]
R_{\mathcal C}
&=
2\cos
\left[
\frac{\arccos(-x)}{3}
\right].
\end{aligned}
\label{eq:appB_radii}
\end{equation}
Expanding these expressions about \(x=0\) gives
\begin{equation}
R_{\mathcal H}
=
\frac{2x}{3}
+
O(x^{3}),
\qquad
R_{\mathcal C}
=
\sqrt{3}
-
\frac{x}{3}
+
O(x^{2}).
\label{eq:appB_radii_expansion}
\end{equation}

The dimensionless surface gravity associated with the black-hole
event horizon is
\begin{equation}
K_{\mathcal H}
=
\frac{
(2R_{\mathcal H}+R_{\mathcal C})
(R_{\mathcal C}-R_{\mathcal H})
}{
6R_{\mathcal H}
}.
\label{eq:appB_KH}
\end{equation}
Using Eq.~(\ref{eq:appB_radii_expansion}), its leading behavior is
\begin{equation}
K_{\mathcal H}
=
\frac{3}{4x}
+
O(1),
\qquad
x\rightarrow0,
\label{eq:appB_KH_limit}
\end{equation}
and therefore
\begin{equation}
K_{\mathcal H}
\rightarrow
\infty.
\label{eq:appB_KH_infinity}
\end{equation}

For any finite dimensionless field frequency \(\varpi\), the
fermionic coefficients
\begin{equation}
\begin{aligned}
\mu_{\mathcal H}
&=
\left[
1+
\exp\left(
-\frac{2\pi\varpi}{K_{\mathcal H}}
\right)
\right]^{-1/2},
\\[6pt]
\nu_{\mathcal H}
&=
\left[
1+
\exp\left(
\frac{2\pi\varpi}{K_{\mathcal H}}
\right)
\right]^{-1/2}
\end{aligned}
\label{eq:appB_H_coefficients}
\end{equation}
then satisfy
\begin{equation}
\mu_{\mathcal H}
=
\nu_{\mathcal H}
=
\frac{1}{\sqrt{2}},
\qquad
x\rightarrow0.
\label{eq:appB_equal_H_coefficients}
\end{equation}
This equality is the origin of the pairwise coincidence of the four
concurrences at \(x=0\).

Consider first
\(C_{\mathrm{GM}}(\rho_{ABC})\) and
\(C_{\mathrm{GM}}(\rho_{LBC})\).
For compactness, define
\begin{equation}
m
=
\frac{1}{\sqrt{2}}.
\label{eq:appB_m}
\end{equation}
Upon setting
\(\mu_{\mathcal H}=\nu_{\mathcal H}=m\)
in Eqs.~(\ref{eq:IV_CGM_ABC}) and
(\ref{eq:IV_CGM_LBC}), the quantity entering the maximization in
both expressions becomes
\begin{align}
\mathcal{F}_{\nu}
={}&
p\alpha\beta
-
\sqrt{
q(1+m^{2})
\left[
p\alpha^{2}\nu_{\mathcal C}^{2}
+
q(1+\nu_{\mathcal C}^{2})
\right]
}
\notag\\
&-
\sqrt{
q
\left[
p\alpha^{2}m^{2}\nu_{\mathcal C}^{2}
+
q(1+m^{2})(1+\nu_{\mathcal C}^{2})
\right]
}
\notag\\
&-
\sqrt{
q(1+\nu_{\mathcal C}^{2})
\left[
p\alpha^{2}m^{2}
+
q(1+m^{2})
\right]
}.
\label{eq:appB_Fnu}
\end{align}
The two concurrences therefore reduce to
\begin{equation}
C_{\mathrm{GM}}(\rho_{ABC})
=
2m\mu_{\mathcal C}
\max\left\{0,\mathcal{F}_{\nu}\right\},
\label{eq:appB_ABC_limit}
\end{equation}
and
\begin{equation}
C_{\mathrm{GM}}(\rho_{LBC})
=
2m\mu_{\mathcal C}
\max\left\{0,\mathcal{F}_{\nu}\right\}.
\label{eq:appB_LBC_limit}
\end{equation}
Hence,
\begin{equation}
C_{\mathrm{GM}}(\rho_{ABC})
=
C_{\mathrm{GM}}(\rho_{LBC}),
\qquad
x\rightarrow0.
\label{eq:appB_ABC_LBC_equal}
\end{equation}

The same procedure applies to
\(C_{\mathrm{GM}}(\rho_{ABR})\) and
\(C_{\mathrm{GM}}(\rho_{LBR})\).
Substituting
\(\mu_{\mathcal H}=\nu_{\mathcal H}=m\)
into Eqs.~(\ref{eq:IV_CGM_ABR}) and
(\ref{eq:IV_CGM_LBR}) yields the common maximization function
\begin{align}
\mathcal{F}_{\mu}
={}&
p\alpha\beta
-
\sqrt{
q(1+m^{2})
\left[
p\alpha^{2}\mu_{\mathcal C}^{2}
+
q(1+\mu_{\mathcal C}^{2})
\right]
}
\notag\\
&-
\sqrt{
q(1+\mu_{\mathcal C}^{2})
\left[
p\alpha^{2}m^{2}
+
q(1+m^{2})
\right]
}
\notag\\
&-
\sqrt{
q
\left[
p\alpha^{2}m^{2}\mu_{\mathcal C}^{2}
+
q(1+m^{2})(1+\mu_{\mathcal C}^{2})
\right]
}.
\label{eq:appB_Fmu}
\end{align}
Consequently,
\begin{equation}
C_{\mathrm{GM}}(\rho_{ABR})
=
2m\nu_{\mathcal C}
\max\left\{0,\mathcal{F}_{\mu}\right\},
\label{eq:appB_ABR_limit}
\end{equation}
while
\begin{equation}
C_{\mathrm{GM}}(\rho_{LBR})
=
2m\nu_{\mathcal C}
\max\left\{0,\mathcal{F}_{\mu}\right\}.
\label{eq:appB_LBR_limit}
\end{equation}
It follows that
\begin{equation}
C_{\mathrm{GM}}(\rho_{ABR})
=
C_{\mathrm{GM}}(\rho_{LBR}),
\qquad
x\rightarrow0.
\label{eq:appB_ABR_LBR_equal}
\end{equation}

Combining
Eqs.~(\ref{eq:appB_ABC_LBC_equal}) and
(\ref{eq:appB_ABR_LBR_equal}), one obtains
\begin{equation}
\begin{aligned}
C_{\mathrm{GM}}(\rho_{ABC})
&=
C_{\mathrm{GM}}(\rho_{LBC}),
\\
C_{\mathrm{GM}}(\rho_{ABR})
&=
C_{\mathrm{GM}}(\rho_{LBR}),
\end{aligned}
\qquad
x\rightarrow0.
\label{eq:appB_pairwise_limits}
\end{equation}

The pairwise coincidence therefore originates from the limiting
relation
\(\mu_{\mathcal H}=\nu_{\mathcal H}=1/\sqrt{2}\).
Once this relation is imposed, the maximization functions and the
corresponding prefactors become identical within each pair.
The result holds for arbitrary allowed values of \(p\) and
\(\alpha\), with
\(\beta=\sqrt{1-\alpha^{2}}\), and for any finite
\(\varpi\).

\bibliography{references}

\end{document}